\documentclass[twocolumn]{aastex631}

\usepackage{amsmath}
\usepackage{graphicx}
\usepackage{float}

\shorttitle{Dust in SN Ibn/Icn 2023xgo}
\shortauthors{Davis et al.}

\newcommand{\pmi}[2]{$_{#1}^{#2}$}

\begin{document}

\title{\textit{JWST} Reveals Large Reservoirs of Dust and Ongoing Circumstellar Interaction in SN Ibn/Icn 2023xgo over a Year Post-Explosion}

\correspondingauthor{Kyle~W.~Davis}
\author[0000-0002-5680-4660]{Kyle~W.~Davis}
\affiliation{Department of Astronomy and Astrophysics, University of California, Santa Cruz, CA 95064, USA}

\author[0000-0002-5748-4558]{Kirsty~Taggart}
\affiliation{Department of Astronomy and Astrophysics, University of California, Santa Cruz, CA 95064, USA}

\author[0000-0002-1481-4676]{Samaporn~Tinyanont}
\affiliation{NARIT, National Astronomical Research Institute of Thailand, 260 Moo 4, Donkaew, Maerim, Chiang Mai, 50180, Thailand}

\author[0000-0002-2445-5275]{Ryan~J.~Foley}
\affiliation{Department of Astronomy and Astrophysics, University of California, Santa Cruz, CA 95064, USA}

\author[0000-0003-3643-839X]{Jeonghee~Rho}\affiliation{SETI Institute, 339 Bernardo Ave., Ste. 200, Mountain View, CA 94043, USA}


\author[0000-0002-4449-9152]{Katie~Auchettl} \affiliation{Department of Astronomy and Astrophysics, University of California, Santa Cruz, CA 95064, USA}\affiliation{OzGrav, School of Physics, The University of Melbourne, Parkville, VIC, Australia}

\author[0000-0002-6886-269X]{Diego~Farias}\affiliation{DARK, Niels Bohr Institute, University of Copenhagen, Jagtvej 128, 2200 Copenhagen, Denmark}\affiliation{INAF, Osservatorio Astronomico di Capodimonte, Salita Moiariello 16, I-80121 Naples, Italy}

\author[0000-0003-2238-1572]{Ori~D.~Fox}\affiliation{Space Telescope Science Institute, Baltimore, MD 21218, USA}

\author[0000-0001-5975-290X]{Joel~Johansson}\affiliation{Department of Physics, Oskar Klein Centre, Stockholm University, SE-106 91, Stockholm, Sweden}

\author[0000-0002-5740-7747]{Charles~D.~Kilpatrick}\affiliation{Center for Interdisciplinary Exploration and Research in Astrophysics (CIERA), Northwestern University, Evanston, IL 60208, USA}

\author[0000-0002-1092-6806]{Kishore~C.~Patra}\affiliation{Department of Astronomy and Astrophysics, University of California, Santa Cruz, CA 95064, USA}

\author[0000-0002-7472-1279]{Craig~Pellegrino}\affiliation{NASA Goddard Space Flight Center, 8800 Greenbelt Road, Greenbelt, MD 20771, USA}

\author[0000-0003-2558-3102]{Enrico~Ramirez-Ruiz}\affiliation{Department of Astronomy and Astrophysics, University of California, Santa Cruz, CA 95064, USA}


\author[0000-0003-4263-2228]{David~A.~Coulter}\affiliation{Department of Physics and Astronomy, The Johns Hopkins University, Baltimore, MD 21218, USA, Space Telescope Science Institute, Baltimore, MD 21218, USA}\affiliation{Space Telescope Science Institute, Baltimore, MD 21218, USA}

\author[0000-0002-7937-6371]{Yize Dong}\affiliation{Center for Astrophysics \textbar{} Harvard \& Smithsonian, 60 Garden Street, Cambridge, MA 02138-1516, USA}

\author[0000-0003-4906-8447]{Alexander~T.~Gagliano}\affiliation{Center for Astrophysics \textbar{} Harvard \& Smithsonian, 60 Garden Street, Cambridge, MA 02138-1516, USA}\affiliation{The NSF AI Institute for Artificial Intelligence and Fundamental Interactions; Physical Superintelligence, Cambridge, MA 02142, USA}

\author[0000-0003-2824-3875]{T.~R.~Geballe}\affiliation{Gemini Observatory/NSF's National Optical-Infrared Astronomy Research Laboratory, 670 N. Aohoku Place, Hilo, HI, 96720, USA}

\author[0000-0002-3934-2644]{Wynn~V.~Jacobson-Gal\'{a}n}\altaffiliation{NASA Hubble Fellow}
\affiliation{Cahill Center for Astrophysics, California Institute of Technology, MC 249-17, 1216 E California Boulevard, Pasadena, CA, 91125, USA}

\author[0009-0000-3122-8321]{Jenna~Karcheski}\altaffiliation{NSF Fellow}\affiliation{Department of Astronomy and Astrophysics, University of California, Santa Cruz, CA 95064, USA}

\author[0009-0005-1871-7856]{Ravjit~Kaur}\affiliation{Department of Astronomy and Astrophysics, University of California, Santa Cruz, CA 95064, USA}

\author[0000-0003-0778-0321]{Ryan~M.~Lau}\affiliation{IPAC, Mailcode 100-22, Caltech, 1200 E. California Blvd., Pasadena, CA 91125, USA}

\author[0000-0001-8385-3727]{Thomas~Moore}\affiliation{Space Telescope Science Institute, Baltimore, MD 21218, USA}

\author[0000-0001-7488-4337]{Seong~Hyun~Park}\affiliation{Department of Physics and Astronomy, Seoul National University, Gwanak-ro 1, Gwanak-gu, Seoul, 08826, South Korea}

\author[0000-0002-4410-5387]{Armin~Rest}\affiliation{Department of Physics and Astronomy, The Johns Hopkins University, Baltimore, MD 21218, USA, Space Telescope Science Institute, Baltimore, MD 21218, USA}\affiliation{Space Telescope Science Institute, Baltimore, MD 21218, USA}

\author[0000-0003-4610-1117]{Tam\'as~Szalai}\affiliation{Department of Experimental Physics, Institute of Physics, University of Szeged, D{\'o}m t{\'e}r 9, 6720 Szeged, Hungary}

\author[0000-0001-5233-6989]{Qinan~Wang}\affiliation{Kavli Institute for Astrophysics and Space Research, Massachusetts Institute of Technology, Cambridge, MA 02139, USA}




\begin{abstract}

We present infrared (IR) photometric and spectroscopic observations of SN~2023xgo, a recent and nearby Type Ibn/Icn supernova (SN~Ibn/Icn) which shows shock interaction with a He/C-rich and H-poor circumstellar medium (CSM). Although interacting SNe are  predicted to produce large amounts of dust, the rarity of SNe~Ibn and Icn has resulted in few opportunities to observe these objects in the IR at late times. Here, we report observations of SN~2023xgo from \textit{JWST} (NIRSpec and MIRI), \textit{WISE}, and Gemini taken out to +377~days post-explosion. At +377~days, the \textit{JWST} spectrum is consistent with both emission from cool ($\sim$300--600~K) silicate dust with  $M \gtrsim 3 \times 10^{-2}$~M$_{\odot}$ at a radius similar to the shock radius ($2.3 \times 10^{16}$~cm), and optically thin carbonaceous dust with $M = 8 \times 10^{-3}$~M$_{\odot}$. We also detect narrow (FWHM\,$ = 520\pm{130}$~km~s$^{-1}$) \ion{He}{1}~$\lambda$2.06~$\mu$m emission at +377~days, indicating that the SN shock continues to encounter material shed from the star to this late epoch. The emission line is blueshifted from the rest frame by 340$\pm{40}$~km~s$^{-1}$. The Gemini and \textit{WISE} observations at $\sim$70--100~days reveal emission from 6.8$\times$10$^{-5}$~M$_{\odot}$ of hot ($\sim$1300~K) dust, which we interpret as a lower limit of the total dust mass at that phase. Molecular gas emission is not detected in any data, though emission line profiles in the optical and NIR taken at $\sim$70~days after explosion show progressively less redshifted emission, attributed to attenuation from dust and suggesting that some dust is rapidly forming interior to the unshocked CSM. The large dust mass and rapid onset of dust formation observed in SN~2023xgo show that the unique physical environments of SNe~Ibn/Icn facilitate substantial dust formation both before and after the SN.

\end{abstract}



\section{Introduction}\label{sec:intro}

Many massive ($\gtrsim$8~M$_\odot$) stars end their lives as energetic and luminous core-collapse supernovae \citep[CCSNe; for a review, see][]{Smartt2009}. The basic observable properties of CCSNe are largely dictated by pre-SN mass loss. Stars which completely shed their hydrogen envelopes prior to core-collapse (whether by strong stellar winds, energetic outbursts, and/or binary interactions) explode as stripped envelope SNe (SESNe) of Type Ib with no apparent hydrogen in their spectra. Even further stripping can remove the entire helium envelope, or any remaining helium may be hidden due to a lack of sufficiently high-energy photons, resulting in a Type Ic SN (SN Ic) that lacks signatures of both hydrogen and helium.


In some instances, a stripped star will explode during, or shortly following, a phase of intense mass loss. In such cases, the SN ejecta will encounter the close-by, dense circumstellar envelope. As the SN ejecta plows into the dense circumstellar medium (CSM), shock interaction efficiently converts the kinetic energy from the explosion to radiation \citep{vanMarle2010} and powers a luminous transient \citep[for a review, see][]{Smith2017}. The spectra of interacting SNe are typically characterized by narrow ($\sim$100--1000~km~s$^{-1}$) spectral emission lines arising from the CSM, along with light curves shaped by the CSM density profile.

Depending on the lines present in their spectra, from which we infer the composition of the CSM, these events are classified as Type IIn \citep[H-rich;][]{Filippenko1989,Schlegel1990}, Type Ibn \citep[He-rich and H-poor;][]{Foley2007, Pastorello2007}, and Type Icn \citep[C/O-rich and H/He-poor;][]{Gal-Yam2022} SNe. These SNe are intrinsically rare, accounting for $\sim$10\%, 2\%, and $<$1\% of the observed local CCSN population, respectively \citep{Perley2020}. More recently, even a single Type Ien SN with C/O/Mg/Ne-poor and S/Si/Ar-rich CSM has been reported \citep{Schulze2025}.


The origins of SNe~Ibn/Icn are particularly confounding. The removal of the entire H (for SNe~Ibn) and He (for SNe~Icn) envelopes sufficiently rapidly such that they are no longer present in the CSM by the time of explosion requires high mass-loss rates that are difficult to reconcile with standard stellar evolution models \citep{Smith2014ARA}. One commonly invoked progenitor channel is the core-collapse of extremely massive ($M_{\mathrm{ZAMS}}\gtrsim40$~M$_{\odot}$) Wolf-Rayet (WR) stars \citep[e.g.,][]{Foley2007, Pastorello2007, Gal-Yam2022}. Towards the ends of their lives, compact WR stars drive H-poor, high-velocity ($\gtrsim$1000~km~s$^{-1}$) stellar winds which create circumstellar environments similar to those observed around SNe~Ibn/Icn. The different surface compositions of WR subtypes, WN being He-rich and WC/WO being He-poor \citep{Schnurr2008}, track to the (observed) compositional differences between SNe Ibn and Icn.

However, several lines of evidence indicate the core-collapse of single WR stars cannot account for all observed SNe~Ibn/Icn. Modeling of the light curves of SNe~Ibn and Icn often recover low ejecta masses ($<$2~M$_{\odot}$) and nucleosynthetic yields ($\lesssim$0.04~M$_{\odot}$~$^{56}$Ni) that are either inconsistent with single WR explosions \citep[e.g.,][]{Pellegrino2022, farias2026}, or require substantial fallback accretion onto a compact remnant. The need for a lower mass progenitor channel is corroborated by analyses of SN~Ibn/Icn host-galaxy environments, which show many 
SNe~Ibn/Icn occur in low-SFR environments and dwarf-like host galaxies inconsistent with the host galaxies of massive WR progenitors \citep{Hosseinzadeh2019, Taggart2021, Pellegrino2022, Davis2023, Warwick2025, Dong2025, Aster2026, Hu2026, Shi2026}.


The necessity of a less massive (than WRs) progenitor channel suggests binary progenitors account for a significant fraction of the SN~Ibn/Icn population \citep[e.g.,][]{Izzard2004}. \citet{Dessart2022} present radiative transfer simulations \citep[applied to observations in][]{Wang2024} of exploding low-mass ($\lesssim$4~M$_{\odot}$) He star binary progenitors which match the bolometric properties of SNe~Ibn/Icn, and \citet{Tsuna2024} show that such systems can replicate the precursor activity observed in some SNe~Ibn. \citet{Ercolino2025} are able to reproduce the CSM properties of some (but not all) SNe~Ibn/Icn using grids of binary evolution models.


Recent work demonstrates that reconstructing the pre-SN mass-loss histories of SNe~Ibn/Icn offers additional insight into their progenitor systems. The previous two direct detections of precursor emission in SNe~Ibn (SNe~2006jc and 2019uo) were consistent with short (lasting $\sim$10~days) and energetic mass ejections \citep{Pastorello2007, Strotjohann2021}. Recently, SN~Ibn~2023fyq showed slowly rising $\gtrsim$3~yr  emission followed by a rapid ($\sim$50~day) brightening preceding the terminal explosion that is difficult for a single massive star to reproduce \citep{Brennan2024, Dong2024}. Radio observations of SN~2023fyq also independently confirm radially varying CSM density profiles extending to at least $10^{16}$~cm, corresponding to mass loss several years prior to explosion  \citep{Baer-Way2025}.

The mass-loss histories of SNe~Ibn beyond $\sim$1~year pre-explosion are challenging to probe. Spectroscopic observations of SNe~Ibn/Icn at $>$100~days (when a shock could breach material shed years before explosion) are extremely limited \citep{Dong2025} and typically require large-aperture telescopes. Similarly, precursor emission has only been detected in the aforementioned 3 nearby SNe~Ibn, although this is set to be revolutionized by the Legacy Survey of Space and Time \citep[LSST;][]{Ivezic2019}. Thus, while scientifically rich, the extended mass-loss histories of SNe~Ibn/Icn remain underexplored.



Owing to their short ($\sim$10~Myr) delay times, CCSNe have long been proposed as a major source of dust in the early Universe \citep[e.g.,][]{Gall2011}. Models of dust formation in the expanding ejecta of SNe~IIP and observations of SN remnants and SN~1987A predict these events can condense substantial dust \citep[$\sim$0.1--1.0~M$_{\odot}$ per SN; e.g.,][]{Todini2001, Gall2011, Sarangi2018}, sufficient to explain the dust reservoirs observed in high-redshift galaxies \citep[e.g.,][]{Marrone2018, Akins2023, Hashimoto2019, Markov2024}.

More recently, \textit{JWST} has enabled direct observations of dust in beyond just the most nearby SNe. Emission from $\sim$10$^{-3}$--10$^{-1}$~M$_{\odot}$ of dust has been detected in late-time ($>$1000~days post-explosion) \textit{JWST} observations of several SNe \citep[e.g.,][]{Shahbandeh2023, Zsiros2024, Sarangi2025, Pearson2025, Clayton2025}, alleviating some tensions between the modest observed dust masses inferred from \textit{Spitzer} observations \citep{Szalai2019}, and the large dust masses predicted in models and observed in SN remnants.

Despite these new \textit{JWST} observations, there are still many gaps in our understanding of high-\textit{z} SN dust production. The basic underlying nucleation processes remain unclear, and it remains unknown whether observed dust mass growth rates reflect a real increase in mass, or arise from optical depth effects \citep{Wesson2015, Dwek2019}. Furthermore, the details of how dust grains form in SN ejecta, grow towards their terminal masses, and are dispersed into the ISM depend on the physical properties and environments of individual SNe themselves \citep[e.g.,][]{Sarangi2015,Marassi2019,Slavin2020,Brooker2022,Sarangi2022b,Liljegren2023,Purushothaman2025}. 

Unique conditions in the early Universe such as lower metallicities, a top-heavy initial mass function (IMF), and differing binary interaction rates are such that SNe, and consequently how they form dust, may not be directly analogous to local SNe~II and SNe~Ibc. In particular, pre-SN mass loss at high \textit{z} may be distinct from that in the local Universe. While low metallicities will suppress wind-driven mass loss, higher stellar masses and binary interaction rates \citep{Sana2025} may increase the amount of SNe that occur within dense CSM.

The presence of dense CSM at the time of explosion is thought to be particularly impactful in shaping dust formation in and around SNe \citep[e.g.,][]{Pozzo2004, Smith2009}. \textit{Spitzer} observations of SNe~IIn show that they are more likely to exhibit luminous dust emission than SNe~IIP, with measured dust masses that can exceed those of SNe~IIP by 1--2 orders of magnitude at the same phase \citep{Fox2011}. Interacting SNe are thought to be especially promising sites for SN dust because, in addition to the SN ejecta, dust can form in (1) the CSM prior to the explosion, and (2) the cold, dense shell (CDS) of gas between the forward and reverse shocks \citep[e.g.,][]{Chugai2004, Sarangi2022a}.



While differentiating the relative contributions of emission from pre-existing dust and newly forming dust in an observed dust SED is non-trivial, recent \textit{JWST} observations of SN~IIn~2005ip and the interacting SN~Ib-pec~2014C at late times ($\gtrsim$10~years post-explosion) provide evidence for $\gtrsim$0.08~M$_{\odot}$ of new dust formed in the CDS since they were last observed by \textit{Spitzer} \citep{Shahbandeh2025, Tinyanont2025}. More broadly, observations of strongly interacting  SNe~II-P and SNe~II-L show evidence for forming dust more promptly ($<$1~year post-explosion) relative to their non-interacting counterparts, although the relative masses of pre-existing circumstellar dust and newly forming dust remain uncertain \citep[e.g.,][]{Andrews2016, Tinyanont2019, Shahbandeh2023, Singh2026}.

The dense and comparatively (relative to SNe~IIn) C/O-rich circumstellar environments of SNe~Ibn/Icn provide conditions that are even more conducive to dust formation. Infrared observations of the prototypical and nearby SN~Ibn~2006jc revealed a thermal excess consistent with $\sim$8$\times 10^{-3}$~M$_{\odot}$ of dust at +230~days \citep{Mattila2008} -- more dust than observed in any other SN by a similar phase -- with up to $\sim$1.5~M$_{\odot}$ of dust expected to condense based on the gas properties \citep{Nozawa2008}. However, both the exact mass and the formation site are uncertain. \citet{Smith2008} argue, based on geometric considerations, that the dust formed in the CDS. Meanwhile, \citet{Nozawa2008} model the SN ejecta and find that rapid cooling can instead enable efficient dust formation in the ejecta. Moreover, \citet{Mattila2008, Sakon2009} model the IR light-curve evolution and assert the bulk of the late-time emission must arise from unshocked circumstellar dust, along with only a modest (albeit detectable and rapidly condensing) amount of newly formed dust in the CDS.

Since the above studies of SN~2006jc, whether or not the remarkable infrared properties of SN~2006jc are ubiquitous among SNe~Ibn has remained unclear. Comparable observational follow-up has been largely impossible for other SNe~Ibn/Icn owing to the intrinsic rarity of the class in the local universe, their blue and rapidly evolving optical emission, and a lack of instrumentation sensitive to wavelengths beyond $\sim$2.5~$\mu$m where the bulk of the dust emission is found.

\citet{Gan2021} analyzed four SNe~Ibn with optical-NIR photometry (out to $K$-band) and identified an IR excess in OGLE-2012-SN-006 consistent with $\sim$10$^{-3}$~M$_{\odot}$ of hot dust, while finding only weak or negligible dust emission in the remaining objects. \citet{Farias2024} directly detect emission from $\sim$10$^{-5}$~M$_{\odot}$ of dust in the peculiar SN~IIn/Ibn~2021foa in spectra covering out to $\sim$2~$\mu$m. For SNe~Icn, \citet{Das2026} recently reported a NIR excess and strong reddening in the light curves of SN~2018erx at +29~days after peak, which they interpret as emission from 10$^{5}$--10$^{6}$~M$_{\odot}$ of pre-existing circumstellar dust. Their observations reach out to ground-based $K$-band, and $\sim$+1~month post-explosion. No IR observations of SNe~Icn extending beyond 2.5~$\mu$m or $\sim$+2~months post-explosion have been reported in literature \citep{Fraser2021, Davis2023, INTEL2026}.

Presented in \citet{Gangopadhyay2025}, SN~2023xgo is a Type Ibn/Icn SN which shows clear evidence of shock interactions with a He/C-rich and H-poor CSM. Located at a distance of $D=54.4$~Mpc, it is the 3rd nearest SN~Ibn (after SNe~2006jc and 2023fyq) and by far the nearest SN~Icn discovered. SN~2023xgo exploded on 09.22 November 2023 UT, and exhibited a rapid rise to a peak optical magnitude of $M_{r}=-17.65$~mag, subluminous relative to most SNe~Ibn  but consistent with several SNe~Icn, followed by a rapid decline of  $\sim$0.14~mag~day$^{-1}$.

The proximity of SN~2023xgo relative to other objects in its class uniquely enabled extensive IR follow-up. The opportunity to study dust formation in this system motivated several successful Director's Discretionary proposals to obtain infrared observations with the Gemini Observatory and \textit{JWST}. Since then, early ($\lesssim$50~days post-explosion), ground-based NIR photometry presented by \citet{Yamanaka2025} revealed a striking IR excess consistent with $\sim$10$^{-5}$~M$_{\odot}$ of dust already present at 17~days post-explosion. Given the prompt appearance of the emission, they argue that the dust is most likely pre-existing in the circumstellar environment.

Taken together, these properties make SN~2023xgo a uniquely powerful laboratory for studying dust formation and late-time circumstellar interaction in SNe~Ibn/Icn. In this paper, we present new NIR and MIR imaging and spectroscopy of the SN~2023xgo obtained with \textit{JWST}, \textit{WISE} and Gemini, along with an analysis of optical emission lines from spectra presented in \cite{Gangopadhyay2025} and \cite{farias2026}. We describe our observations in Section~\ref{sec:obs}. We then fit the \textit{JWST} spectra with dust models and attempt to constrain the dust location and ongoing circumstellar interaction in Section~\ref{sec:analysis}. We discuss the possible dust heating source, constraints on the progenitor for SN~2023xgo, and implications for early-universe dust in Section~\ref{sec:discussion}. Lastly, we summarize our results and conclude in Section~\ref{sec:conclusion}.

We adopt a distance to SN~2023xgo of $d=54.4$~Mpc derived from the redshift $z=0.01325$ from \citet{Gangopadhyay2025}. All dates mentioned in the text are given as UT dates or MJD, and all phases are in reference to the inferred explosion time of 09.22 November 2023 UT (MJD 60257.22) from \citet{Gangopadhyay2025} unless otherwise explicitly stated. We also adopt their combined host-galaxy and Milky Way extinction estimate of $A_V=0.49$~mag.

\section{Observations and Data Reduction} \label{sec:obs}
\begin{figure}
\begin{center}
\includegraphics[width=0.45\textwidth]{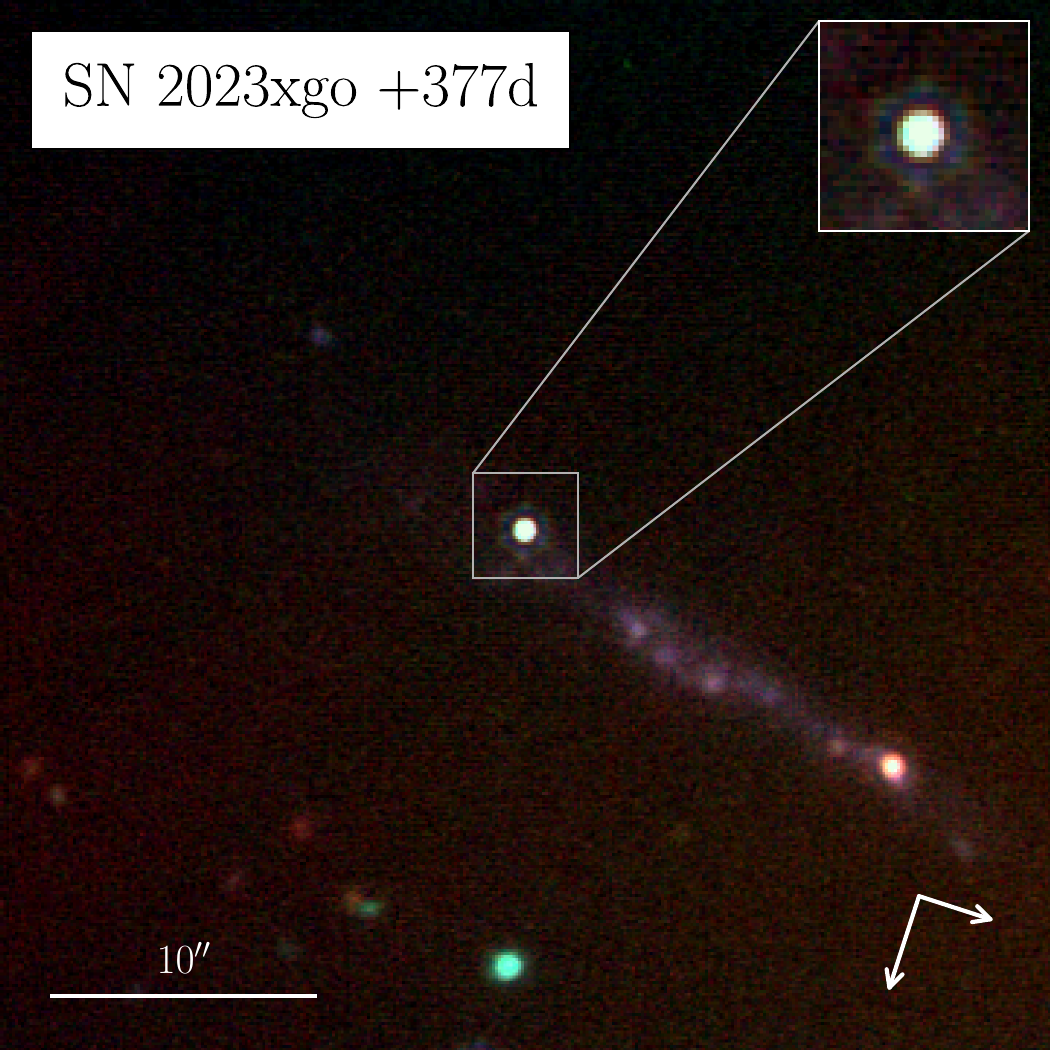}
    \caption{Composite color cutout image of SN~2023xgo at +377~days post-explosion from MIRI images. F1280W, F1500W, F1800W, and F2100W filters are assigned to each color channel. We show a field-of-view of 40-by-40$''$, with a 2-by-2$''$ cutout of the SN inset in the upper-right. SN~2023xgo is clearly detected. The host galaxy is seen trailing to the lower-right of the SN, and we expect minimal emission from the host at the SN location.}
    \label{fig:finder}
\end{center}
\end{figure}

\begin{figure*}
\centering
\includegraphics[width=\textwidth]{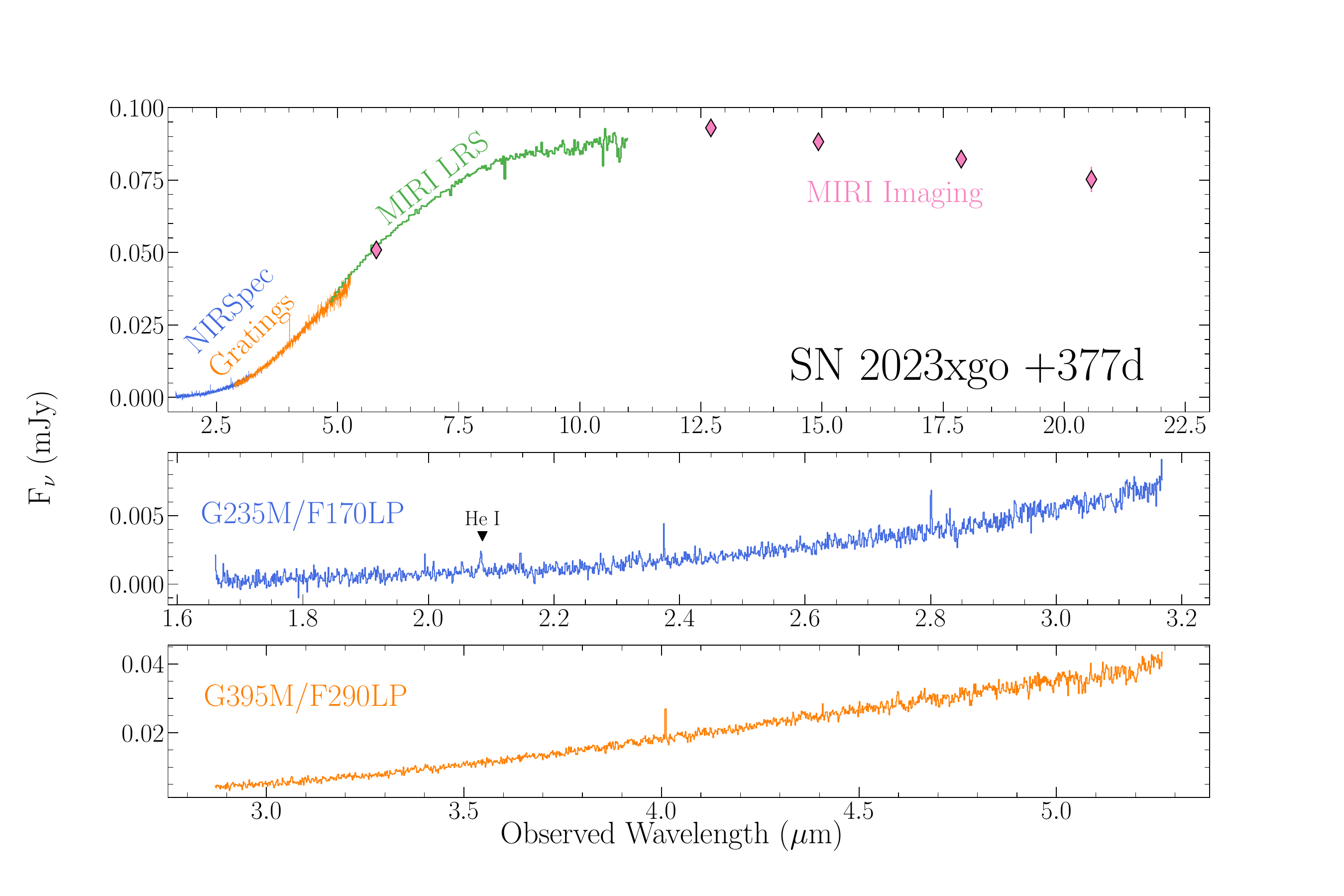}
\caption{Full SED of 2023xgo from \textit{JWST} observations at +377 days post-explosion. The NIRSpec G235M and G395M spectra are shown in blue and orange respectively and additionally in the two lower panels. MIRI LRS data are shown in green. MIRI imaging data are shown in pink diamonds. The flux calibration is taken directly from the data reduction pipeline, and we show the data without any other normalizations.}

\label{fig:JWST_spectrum}

\end{figure*}

SN\,2023xgo was observed with {\it JWST} as part of Cycle 3 program DD-6838 (PI Davis) on 25 November 2024 (377 rest-frame days post-explosion) using the Near-InfraRed Spectrograph (NIRSpec) in the integral-field unit (IFU) mode \citep{Jakobsen2022, Boker2023}, and the Mid-InfraRed Instrument (MIRI) in the Low-Resolution Spectrometer (LRS; \citealp{Kendrew2015}) and Imaging \citep{Bouchet2015, Ressler2015, Rieke2015} modes. We show a finder chart from the MIRI imaging in Fig.~\ref{fig:finder}, and the full \textit{JWST} SED in Fig.~\ref{fig:JWST_spectrum}. All \textit{JWST} data used in this paper can be found in MAST: \href{https://archive.stsci.edu/doi/resolve/resolve.html?doi=10.17909/6s27-mc97}{10.17909/6s27-mc97}.

SN\,2023xgo was also observed with the Gemini Near InfraRed Spectrograph \citep[GNIRS;][]{elias2006a,elias2006b} on the Gemini North telescope as part of programs GN-2023B-Q-225 (PI Rho) and GN-2023B-DD-108 (PI Davis) at 6, 71 and 102 rest-frame days post-explosion. 

Lastly, we report observations from the NEOWISE-reactivation mission taken between 99 and 104 days post-explosion. We describe our observing strategies and data reduction below. Tabulated logs of our observations are provided in the Appendix.

\subsection{\textit{JWST} NIRSpec IFU}
The NIRSpec IFU observations were obtained using the G235M/F170LP and G395/F290LP disperser/filter combinations. This setup achieves a nominal resolving power of $R\approx1000$ over a continuous observer-frame wavelength range of 1.66-5.10~$\mu$m. All observations used a \texttt{4-POINT-DITHER} dither pattern and the \texttt{NRSIRS2RAPID} readout pattern. At each dither position, we take 3 (2) integrations-per-exposure, with 5 (4) groups-per-integration resulting in a total exposure time of 4551 (2451)~s in the G235M/F170LP (G395M/F290LP) configuration.

We use the \textit{JWST} data-reduction pipeline v1.17.1 to process the raw data with the Calibration Reference Data System (CRDS) v11.17.25. We use all default reduction parameters to construct the calibrated 3-D data cube. When viewing the extracted 1-D spectrum on MAST using default extraction parameters, there are clear oversubtraction issues present in the G235M/F170LP mode, as well as known cube-building artifacts. Thus, we perform our own spectral extraction by performing aperture photometry along each slice of the calibrated stage 3 data cube using {\tt photutils}. We use a circular aperture with a radius of 2 pixels (0.2") centered on the SN, and apply a wavelength-dependent aperture correction determined from synthetic PSF correction curves from {\tt STPhot}. We test two methods of determining the local background flux: an annulus with inner and outer radii of 3 and 4~pixels respectively, and a 5th-degree polynomial fit to the background. Both methods yield nearly identical extractions, and so we proceed with the annular background subtraction due to a slightly higher signal-to-noise ratio. For further details on the extraction and background estimation, see \citet{Patra2025}.



\subsection{\textit{JWST} MIRI LRS}
The MIRI LRS observation was obtained using LRS slit mode. This setup covers $\sim$5-14~$\mu$m (we recover useful data out to 10~$\mu$m, beyond which calibration artifacts and low S/N drastically bias the continuum shape) with resolving power ranging from $R\approx50$ to $R\approx160$ at the blue and red ends respectively. We use the default \texttt{ALONG SLIT NOD} dither pattern, with two dither points 1.9" apart on the slit. We use the default \texttt{FASTR1} read pattern, and 1 exposure with 24 integrations and 26 groups-per-integration for a total exposure time of 3591~s. The calibrated and extracted 1-D spectrum was obtained from the Mikulski Archive for Space Telescopes (MAST) using the automatic \textit{JWST} data reduction pipeline (calibration pipeline v1.17.1 at the time of retrieval).

\subsection{\textit{JWST} MIRI Imaging}
We obtained imaging of SN~2023xgo with \textit{JWST}/MIRI on 2024 November 25 (377 rest-frame days post-explosion) in F1280W, F1500W, F1800W, and F2100W filters. We used the \texttt{FULL} array, the \texttt{FASTR1} readout pattern, and a 4-point dither pattern. We also obtained one image in F560W as a target acquisition image for our MIRI LRS observations using the \texttt{FASTGRPAVG8} readout pattern.

Because of the complex MIRI PSF at longer wavelengths and relatively clean background, we choose to perform aperture photometry to estimate the flux from SN~2023xgo in our images. Although the SN sits atop its host galaxy, the expected flux contribution at the location of the SN is quite low ($<1\%$ of the SN flux). At longer wavelengths, the background is dominated by the instrumental background.  

We perform aperture photometry using {\tt space\_phot} \citep{Pierel2024}. Flux is extracted from an aperture centered on the SN location at each individual dither position and averaged for final value. We set aperture widths at between 60--70\% encircled energy, and apply the corresponding aperture corrections supplied by calibration pipeline. Background annuli are set to their default values determined by the aperture size. We find that the scatter in the measured flux at each dither location is larger than the formal uncertainty from each extraction, and so adopt this as a more representative value of the true uncertainty. Extracted fluxes and exposure times are reported in Table~\ref{tab:photlog}.

\subsection{GNIRS NIR Spectroscopy}
Three epochs of NIR spectroscopy were obtained with GNIRS aboard the Gemini North telescope using the cross-dispersed mode at +6, +71, and +102~days post-explosion.  The observations took place on 2023 November 15, 2024 January 20, and February 20.  We use the 32~line~mm$^{-1}$ grating and 0.45~arcsec slit, which achieves simultaneous wavelength coverage from $\sim$0.8--2.5~$\mu$m with a nominal resolving power of $R\approx 1200$. We extract and perform relative flux calibration using {\tt PypeIt v1.18.1}, and apply telluric corrections using {\tt xtellcor} and observations of a nearby telluric standard taken immediately prior-to or after the science exposures.

\subsection{NEOWISE Imaging}
We also present photometric observations from the NEOWISE-reactivation mission \citep[NEOWISE-R;][]{mainzer2011, mainzer2014}. Single exposures in the \textit{W1} and \textit{W2} bands (with central wavelengths of 3.4 and 4.6~$\mu$m, respectively) were collected from the NASA/IPAC Infrared Science Archive (IRSA) and coadded using the ICORE service \citep{masci2013}. We stack images from 29 visits between MJD 60357.87 and 60362.06 (99.3 and 103.5 rest-frame days post-explosion, respectively). Reference images were created by co-adding pre-SN images, which were subtracted from the SN images. 
We perform aperture photometry on the subtracted images using a standard 8.25\arcsec radius aperture, and zero magnitude fluxes for the WISE \textit{W1} and \textit{W2} bands of $F_{\nu,0}^{W1} = 306.7$ Jy and $F_{\nu,0}^{W2} = 170.7$ Jy \citep{Wright2010}.

\section{Dust Emission and Circumstellar Environment}\label{sec:analysis}

We present analysis of the IR observations listed above, as well as the optical spectra of SN~2023xgo presented in \citet{Gangopadhyay2025} and \citet{farias2026}. We model the observed SEDs with dust emission in Sections~\ref{sec:dust}~and~\ref{sec:earlydust}, and constrain the dust location within the system in Section~\ref{sec:dustloc}. Lastly, we discuss the detection of ongoing circumstellar interaction in Section~\ref{sec:interaction}.

\subsection{{\it JWST} Spectrum Fitting and Dust Parameter Estimation}\label{sec:dust}
The SED of SN~2023xgo at +377~days can be qualitatively described as a smooth, cool dust continuum with no apparent emission from the SN or hot dust grains emitting near their vaporization temperatures ($\gtrsim$1500~K; see Section~\ref{sec:earlydust}). We fit the IR spectrum of SN\,2023xgo with several analytical dust models to constrain the dust temperature, mass, composition, and radius.

\begin{figure*}
\centering
\includegraphics[width=\textwidth]{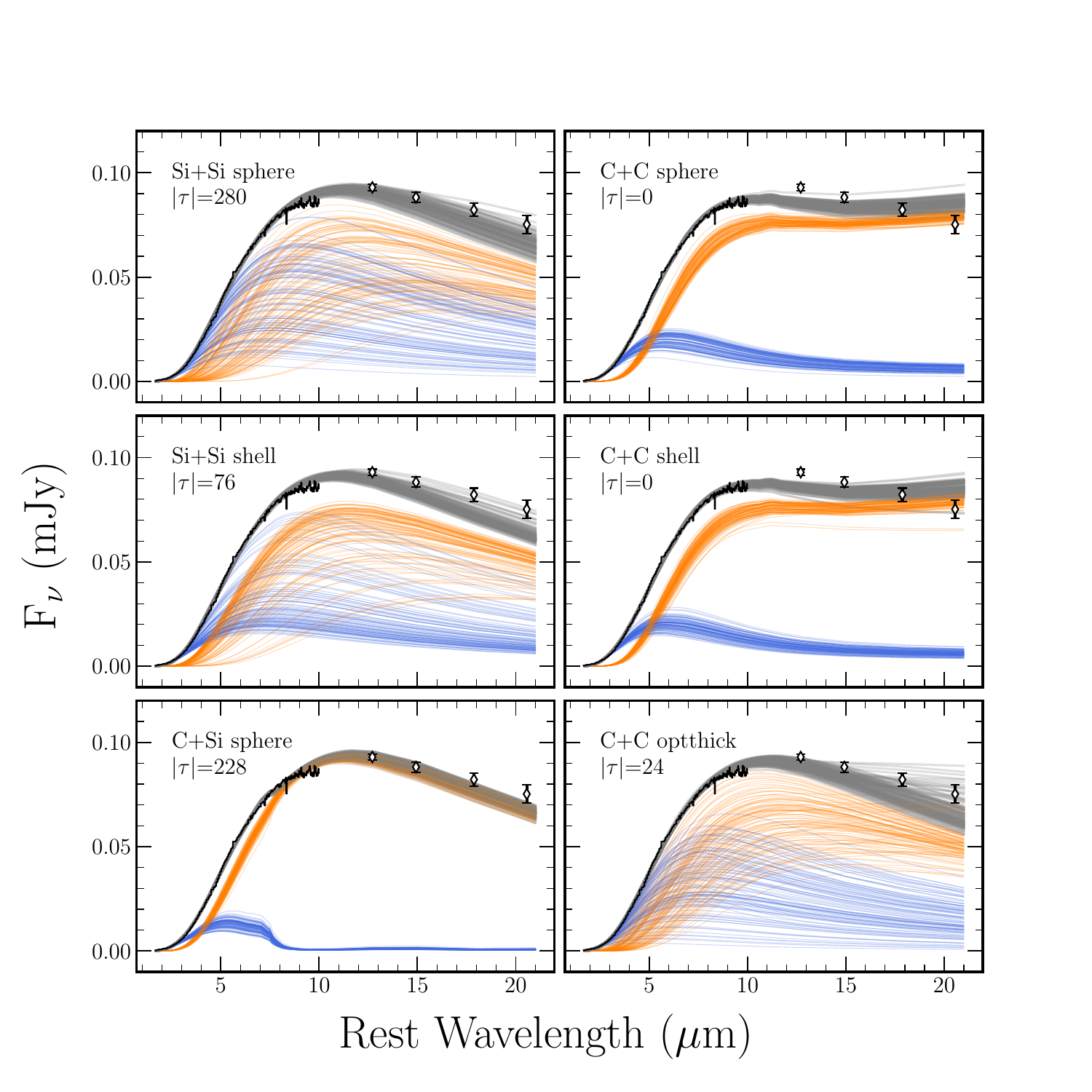}
\vspace{-24pt}
\caption{SED fits for six dust emission models used in our analysis. Results are presented in Table~\ref{tab:dustmodels}. Model details including compositions for the two dust components (C for carbonaceous and Si for silicate) and the assumed geometry (uniform sphere, spherical shell, and optically thick sphere) are encoded into the label shown in the body of each subplot. We plot the input SED in black, as well as 150 draws from each posterior distribution in gray. The drawn warm components are plotted in blue, while the corresponding cool components are plotted in orange. Optical depth values shown are averaged both over wavelength and the 150 posterior draws.
\label{fig:dustmodels}}
\end{figure*}

We test a total of six models, shown in Fig.~\ref{fig:dustmodels}. In each of our models, we assume there to be emission from two distinct components of dust which can differ in mass, temperature, and composition. Both dust components are assumed to be uniformly distributed within the same geometry. The first four models in Fig.~\ref{fig:dustmodels} consist of purely carbonaceous or purely silicate compositions in uniform spherical (denoted by ``\texttt{sphere}" in the model name) or spherical shell (denoted by ``\texttt{shell}") geometries. A fifth model utilizes a mixed composition in a spherical configuration. Lastly, a sixth model consists of purely carbonaceous dust in a spherical geometry wherein we restrict the radius such that the dust is close to optically thick (denoted by ``\texttt{optthick}").

The set of equations presented by \cite{Shahbandeh2023} is employed. The observed dust flux is given by
\begin{equation}
    F_{\rm dust}(\lambda) = \frac{B(\lambda, T_{\rm dust}) \kappa(\lambda) M_{\rm dust} P_{\rm esc}(\tau) }{d^2},
\end{equation}
where $B$ is the Planck function, $d$ is the distance to the SN from \citet{Gangopadhyay2025}, and $\kappa$ is the dust opacity from \citet{Draine1984} and \citet{Laor1993} assuming a grain size of $a = 0.1 \ \mu$m.
In the IR, $\lambda \gg a$, and the spectral shape is insensitive to the grain size \citep[e.g.,][]{Fox2010, Sarangi2022a}.  

$P_{\rm esc}(\tau)$ and the optical depth $\tau$ itself depend on the geometry of the dust. Because the dust in SN~2023xgo may be in any/all of the ejecta, CDS, and CSM, we model the SED with both a uniform sphere, and a spherical shell. 

In the uniform sphere case, the escape probability $P_{\rm esc, sphere}(\tau)$ is given by \citet{Cox1969} as 
\begin{equation}\label{eq:Pesc_sphere}
P_{\rm esc, sphere}(\tau) = \frac{3}{4\tau} \left[1- \frac{1}{2\tau^2} +
                                        \left(\frac{1}{\tau}+\frac{1}{2\tau^2} \right) e^{-2\tau}
                                    \right].
\end{equation} The optical depth, assuming that the dust is evenly distributed, is 
\begin{equation}\label{eq:tau}
    \tau(\lambda) = \rho(t) R \kappa(\lambda) = \frac{3}{4} \frac{M_{\rm dust} }{\pi R^2} \kappa(\lambda).
\end{equation}

For a spherical shell, the photon escape probability is modified due to the interior cavity. We refer to the equations for $P_{\rm esc}$ and $\tau$ derived in \cite{Dwek2024} and applied in e.g. \cite{Pearson2025} wherein 

\begin{equation}\label{eq:Pesc_shell}
    P_\mathrm{esc,shell} = \frac{1}{2\tau} \left[\frac{f(u, \tau)}{f_0(u)}\right],
\end{equation}
where
\begin{equation}
    \begin{split}
    f(u, \tau) = & \int_0^{x_c} \left[ 1 - e^{-2\tau x} \right] x \, dx \\
    &+ \int_{x_c}^1 \left[ 1 - e^{-2\tau x \left( 1 - \sqrt{1 - \frac{1 - u^2}{x^2}} \right)} \right] x \, dx,
    \end{split}
\end{equation}
and
\begin{equation}
    \begin{split}
    f_0(u) = & \int_0^{x_c} x^2 \, dx \\
    &+ \int_{x_c}^1 \left[ 1 - \sqrt{1 - \frac{1 - u^2}{x^2}} \right] x^2 \, dx,
    \end{split}
\end{equation}
where $x_c = \sqrt{1-u^2}$, $u = \frac{R_\mathrm{in}}{R_\mathrm{out}}$, and $R_\mathrm{in}$ and $R_\mathrm{out}$ are the inner and outer radii of the shell, respectively. The optical depth is also given as
\begin{equation}\label{eq:tau_shell}
\tau = \frac{3 \kappa R_\mathrm{out}}{4 \pi (R_\mathrm{out}^3-R_\mathrm{in}^3)}(M_{dust}).
\end{equation}

We leave the radius (radii in the shell case) as a free parameter in our fits due to the uncertain formation region of the dust.

Prior to fitting the data, we resample the NIRSpec spectra into bins of 0.05~$\mu$m (similar to the native spectral bin size of MIRI LRS) using the flux-conserving approach built into {\tt specutils}. Because we simultaneously fit non-overlapping spectroscopy and imaging, we modify our likelihood function such that all of the spectroscopic data points in totality have equal weight to the imaging data points. 

To sample the posterior distributions of our fit parameters, we perform a Markov chain Monte Carlo (MCMC) fit using {\tt emcee} \citep{emcee}. We run 12$\times$N walkers through 3000 iterations where N is the number of free parameters in our fit (5 in spherical models, and 6 in spherical shell models). Convergence is assessed using the integrated autocorrelation time, and an initial burn-in phase is discarded.

Full details of our priors can be found in the Appendix. Results for all 6 models can be found in Table~\ref{tab:dustmodels} and Fig.~\ref{fig:dustmodels}. We find that the inner radius for our shell fits cannot be meaningfully constrained, and thus focus our analysis on the uniform sphere models. We provide corner plots for our pure silicate and carbonaceous spherical dust models in the Appendix, whereas corner plots for the other models will be made available in online supplemental materials.

\begin{table*}
\centering
 \caption{Dust Modeling Results} \label{tab:dustmodels}
 \hspace*{-1.6cm}
 \begin{tabular}{ l c c c c c c}
    \hline
     Model & $M_{1}$ & $M_{2}$ & $T_{1}$ & $T_{2}$ & $R_{\rm in}$ & $R_{\rm out}$ \\
      & [M$_{\odot}$] & [M$_{\odot}$] & [K] & [K] & [cm] & [cm] \\
     \hline
     Si+Si sphere\tablenotemark{a} & $>$2.6$\times$10$^{-2}$ & $>$2.5$\times$10$^{-3}$ & 385\pmi{-51}{+42} & 608\pmi{-36}{+61} & -- & 2.3\pmi{-0.2}{+0.3}$\times$10$^{16}$\\
     
     C+C sphere\tablenotemark{b} & 7.2\pmi{-0.4}{+0.5}$\times$10$^{-3}$ & 2.0\pmi{-0.5}{+0.6}$\times$10$^{-4}$ & 315\pmi{-5}{+6} & 526\pmi{-15}{+19} & -- & $>$5.5$\times$10$^{16}$\\
     
     Si+Si\_shell\tablenotemark{a} & $>$5.9$\times$10$^{-3}$& $>$1.4$\times$10$^{-3}$ & 428\pmi{-31}{+15} & 660\pmi{-47}{+42} & unconstrained & 2.1\pmi{-0.1}{+0.1}$\times$10$^{16}$\\
     
     C+C shell\tablenotemark{b} & 7.8\pmi{-0.5}{+0.5}$\times$10$^{-3}$ & 2.0\pmi{-0.5}{+0.5}$\times$10$^{-4}$ & 315\pmi{-6}{+6} & 525\pmi{-16}{+22} & unconstrained & $>$5.5$\times$10$^{16}$\\
     
     C+Si sphere\tablenotemark{a}& $>$1.7$\times$10$^{-2}$ & $>$3.4$\times$10$^{-4}$ & 442\pmi{-9}{+9} & 613\pmi{-20}{+25} & -- & 2.3\pmi{-0.1}{+0.1}$\times$10$^{16}$ \\
     
     C+C optthick\tablenotemark{a}& $>$7.3$\times$10$^{-3}$ & $>$2.5$\times$10$^{-4}$ & 407\pmi{-81}{+39} & 634\pmi{-51}{+81} & -- & 2.3\pmi{-0.2}{+0.9}$\times$10$^{16}$ \\
     
    \hline

 \end{tabular}

\tablenotetext{a}{Emission is optically thick. Masses reported are 2.5th-percentile lower limits.}
\tablenotetext{b}{Emission is optically thin. Dust radius lower bound is set by the blackbody radius.}

\end{table*}

We find that the IR SED of SN~2023xgo is consistent with both the pure silicate and pure carbonaceous uniform sphere models, between which we cannot formally distinguish a better fit. For the silicate model, the radius is well constrained to 2.3$\times$10$^{16}$~cm, at which point the dust becomes optically thick (as can be seen by the lack of strong 10/18~$\mu$m silicate dust features). In this case, as stated in \citet{Dwek2019} and \citet{Shahbandeh2023}, an arbitrarily high dust mass can fit the data as emission from additional dust does not escape to the observer. Thus, the ``best-fit" mass extracted from our posteriors (which are non-Gaussian) does not meaningfully represent the true dust mass. We conservatively use the lower 2.5th-percentile value to ascertain a reasonable lower limit of the dust mass which comes out to total 2.8$\times$10$^{-2}$~M$_{\odot}$. We note that in this case, the exact mass limit is dependent on our prior. 

For the carbonaceous sphere model, the radius does not converge to a single value, and extends to seemingly arbitrarily high values. This indicates that the dust is mostly optically thin, in which case the total measured mass of 8.1$\times$10$^{-3}$~M$_{\odot}$ is representative of the true total dust mass (barring colder dust hidden at higher wavelengths). While the mass does differ by a factor of $\gtrsim$3 between models, even the lower estimate from the carbonaceous dust is among the highest dust masses ever seen in SNe $\sim$1~year post-explosion. 

The temperatures of the two dust components are similar across models. We measure $T_{\mathrm{C, warm}}= 526_{-15}^{+19}$~K and $T_{\mathrm{C, cold}}= 315_{-5}^{+6}$~K for the carbonaceous dust, and $T_{\mathrm{Sil, warm}}= 608_{-36}^{+61}$~K and $T_{\mathrm{Sil, cold}}= 385_{-51}^{+42}$~K for silicates.

 
%

The inferred dust radius differs drastically between the models. The silicate dust is optically thick, and therefore the radius is constrained at 2.3$\times$10$^{16}$~cm. On the other hand, the carbonaceous dust is optically thin, in which case the radius can be arbitrarily large. We further discuss the modeled dust radius alongside other constraints in Section~\ref{sec:dustloc}.



%

\subsection{Early Evidence for Dust}\label{sec:earlydust}

\begin{figure}
\includegraphics[width=0.48\textwidth]{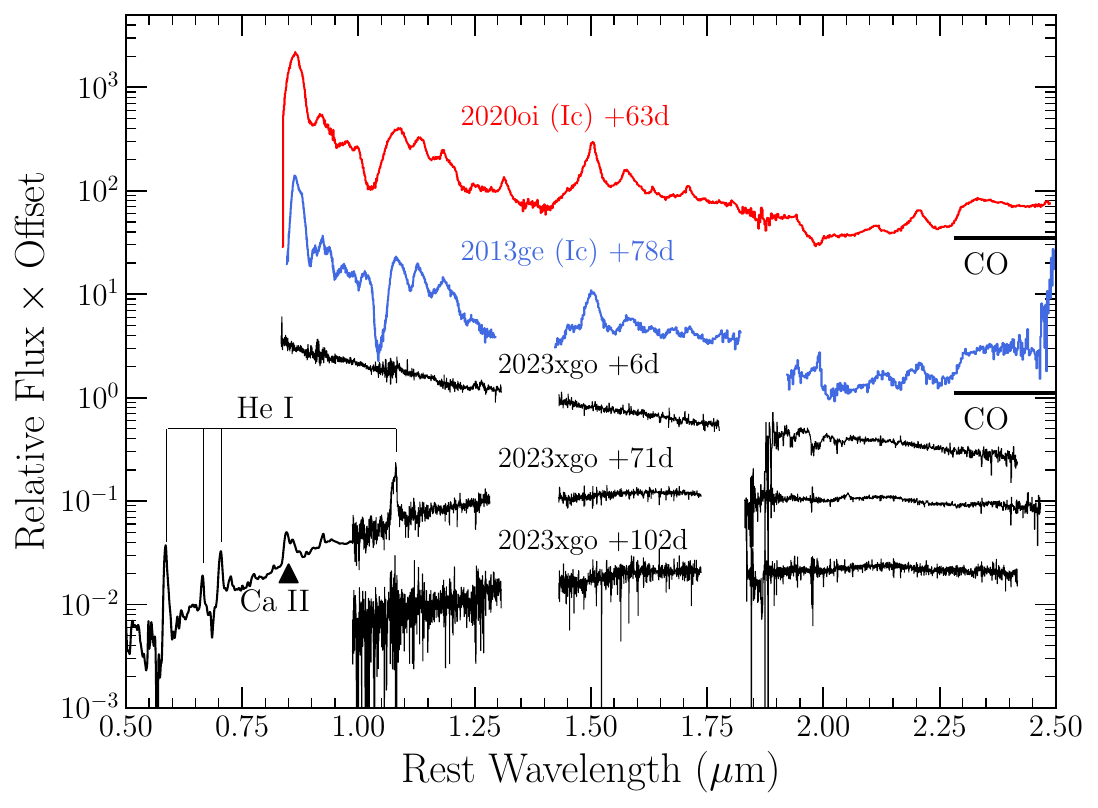}
    \caption{Ground-based NIR spectra of SN~2023xgo compared with NIR spectra of SNe~Ic at $\sim$+70~days \citep[from][]{Shahbandeh2022}. We also show the optical spectrum at +67~days from \citet{farias2026}. SN~2023xgo spectra are dominated by a $\sim$1200~K dust continuum superimposed with He emission lines by +71~days. Unlike SN~2013ge and SN~2020oi \citep{Rho2021}, CO is not detected in SN~2023xgo.}
    \label{fig:nir_compare_early}
\end{figure}

We present NIR spectra of SN~2023xgo at +71~and~+102~days post-explosion in Fig.~\ref{fig:nir_compare_early}. By +71~days, the spectra show a strong NIR excess which we interpret as emission from hot dust. The spectra are dominated by a $\sim$1200~K dust continuum, but also show emission from \ion{He}{1}~$\lambda$1.08~$\mu$m (strongly detected at +71~days, marginally detected at +102~days), \ion{He}{1}~$\lambda$2.06$~\mu$m (detected at +71~days, non-detection at +102~days) indicating that CSM interaction persists. Hot dust emission in SN~2023xgo was first reported in photometric observations by \cite{Yamanaka2025}, who showed a NIR excess was present by +15~days post-explosion.

We model the emission at +102~days using the GNIRS spectrum and \textit{WISE} photometry to constrain the dust mass and temperature. Unlike our \textit{JWST} spectra, the GNIRS data are not absolute-flux calibrated, and therefore cannot themselves constrain the dust mass. We first fit the GNIRS spectrum to obtain a dust temperature. We then scale the model dust mass to match the \textit{WISE} photometry assuming emission from a single dust component at the temperature derived from the spectrum. We assume optically thin emission as the data cannot constrain the radius (and this yields a lower limit of the dust mass).

\begin{figure}
\includegraphics[width=0.46\textwidth]{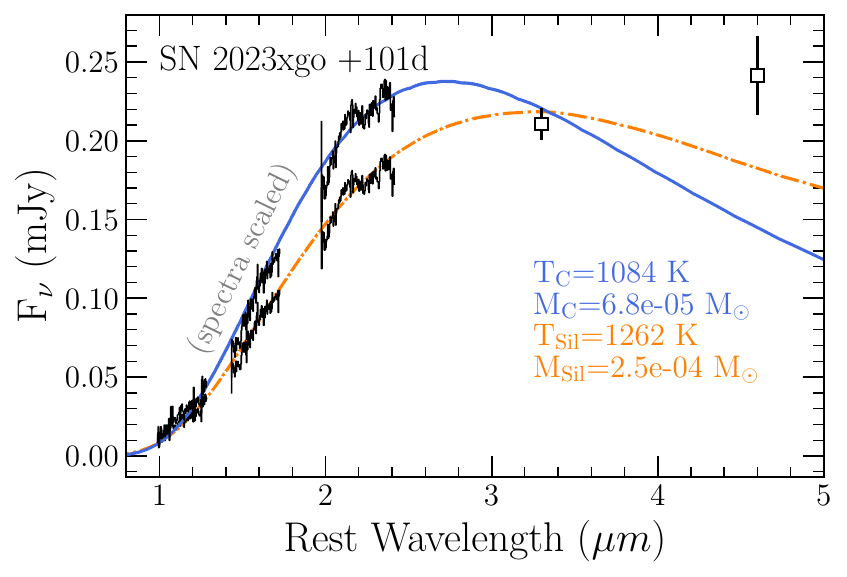}
    \caption{Best-fit optically thin dust models to Gemini and \textit{WISE} observations of SN~2023xgo at +101~days post-explosion. Carbonaceous dust model is shown in blue, while silicate dust model is shown in orange. We obtain the dust temperature by fitting the GNIRS spectrum independent of the \textit{WISE} photometry. We then determine a dust mass by scaling
    the dust SED to the \textit{WISE} observations (with the assumption that the dust emits exclusively at the fitted temperature). The fitted dust masses represent a lower limit to the true dust mass.
    }
    \label{fig:nir_dust}
\end{figure}

We perform a least-squares minimization fit using \texttt{curve\_fit} from \texttt{scipy.optimize} for both silicate and carbonaceous dust (Fig.~\ref{fig:nir_dust}). The differences in derived mass are small: M$_{\mathrm{C}}$$=$6.8$\times$10$^{-5}$~M$_{\odot}$ and M$_{\mathrm{Sil}}$$=$2.5$\times$10$^{-4}$~M$_{\odot}$. We adopt the lower carbonaceous dust mass of 6.8$\times$10$^{-5}$~M$_{\odot}$ as our lower limit.

\citet{Yamanaka2025} report a constant dust temperature of $\sim$1600~K throughout their observations, which is above the vaporization temperature of silicate dust ($\sim$1300~K), but achievable for carbonaceous dust. However, we measure dust temperatures $<$1300~K at +71~days in our NIR spectrum, which is below the vaporization temperature for silicate dust grains. The discrepancy could be due to strong He emission in their $J$-band photometry. Therefore, the composition of the early dust is uncertain. This hot dust emission is not seen in the \textit{JWST} spectrum (which rules out a mass of dust emitting at 1200~K dust above $\sim$10$^{-6}$~M$_{\odot}$). 

We note the obvious limitation in assuming a single dust component when in reality (as seen in the \textit{JWST} spectrum), there may be multiple. However, because we measure the temperature from NIR observations, our methodology yields a reasonable lower limit of the dust mass. Luminous emission from hotter dust is ruled out by the NIR spectra, whereas colder dust would require a higher mass to produce the luminosity in the \textit{WISE} bands.

\begin{figure}
\includegraphics[width=0.48\textwidth]{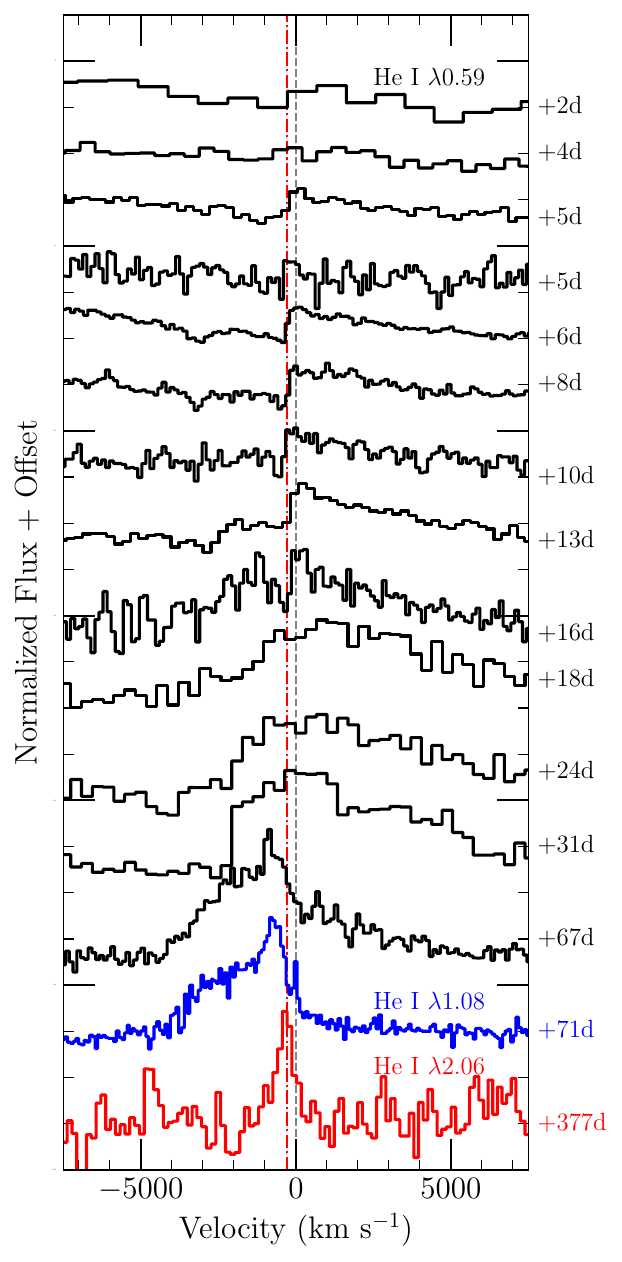}
    \caption{
    Evolution of \ion{He}{1} line profiles in velocity space relative to the rest frame of the host galaxy. The gray vertical dashed line shows 0~km~s$^{-1}$. By +67~days, the emission is clearly blueshifted, most likely due to optically thick dust forming interior to the emitting region \citep[e.g.][]{Smith20082006jc}. By this epoch, the emission lines in both the optical and NIR show an intermediate-width component extending to $\sim-$5000~km~s$^{-1}$, and a narrow peak centered at $\sim-$700~km~s$^{-1}$. By +377~days, the center of the narrow component has receded to $-$340~km~s$^{-1}$ (plotted in a red vertical dash-dotted line). Optical spectra are taken from \citet{Gangopadhyay2025} and \citet{farias2026}.
    }
    \label{fig:attenuate}
\end{figure}

We show the evolution of the helium line profiles using optical spectra presented in \citet{farias2026} and our NIR spectrum in Fig.~\ref{fig:attenuate}. By $\sim$+70~days, both the \ion{He}{1}~$\lambda$0.59~$\mu$m and \ion{He}{1}~$\lambda$1.08~$\mu$m line profiles consist of emission from an intermediate-width ($FWHM\approx 2290$~km~s$^{-1}$, either ejecta or CDS) and narrow ($FWHM\approx 1060$~km~s$^{-1}$) component. Both components appear strongly blueshifted, with the narrow component peaking at $-$650~km~s$^{-1}$ relative to the rest frame, and the flux dropping precipitously red-ward of 0~km~s$^{-1}$.

The increasingly blueshifted line profiles of SN~2023xgo are similar to that observed in SN~2006jc \citep{Foley2007} and SN~2010jl \citep{Gall2014}, and arises from new dust forming interior to the line-emitting region and thus preferentially absorbing photons from the receding (redshifted) side of the system. The line profiles imply a complex scenario where both pre-existing circumstellar dust (reported to have been observed 15~days post-explosion, before any apparent attenuation of the line profiles) and new dust forming in the SN ejecta/CDS (evidenced by the attenuation of the broad emission line component) are present in unknown relative quantities.

\begin{figure}
\includegraphics[width=0.49\textwidth]{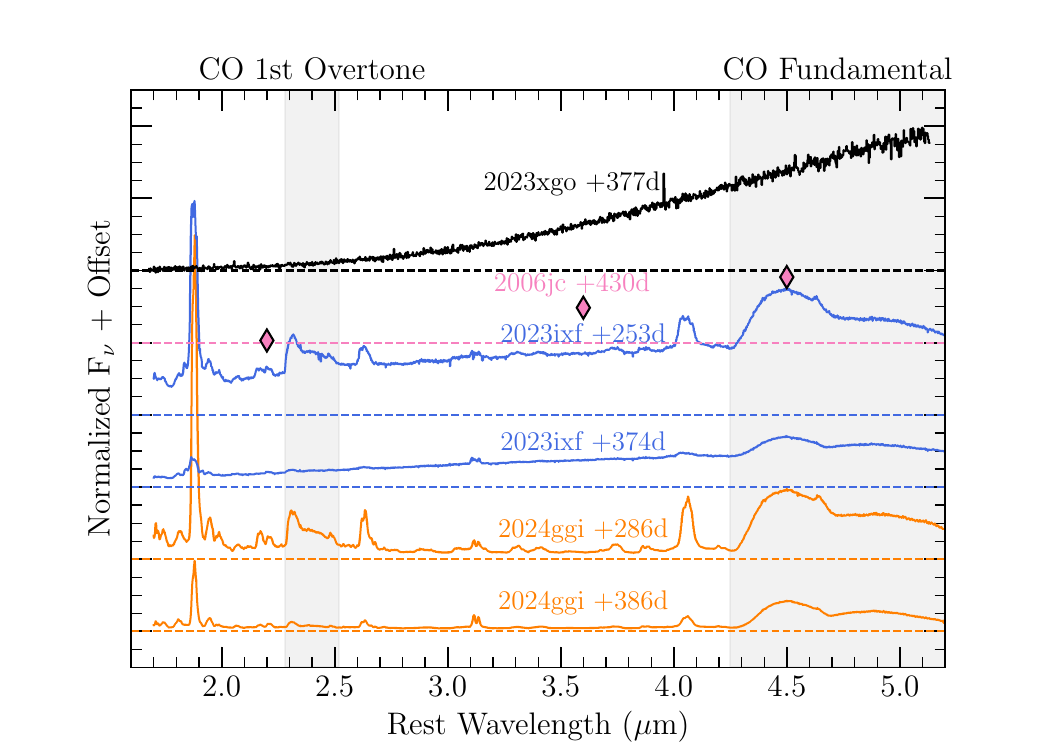}
    \caption{Comparison of $\sim$+1~year NIR spectrum of SN~2023xgo with that of other CCSNe \citep[SN~2006jc, SN~2023ixf, SN~2024ggi][]{Mattila2008, Medler2025, Mera2026}. The fluxes are normalized to a common distance, then offset for visual clarity (dashed lines show $F_{\nu}=0$ for each spectrum). CO fundamental and first overtone emission bands are shown in the grayed-out regions. Unlike the spectra of SN~2023ixf and SN~2024ggi which continue to show luminous CO emission, especially in the fundamental band, CO emission is not detected in the spectrum of SN~2023xgo.
    }
    \label{fig:nir_compare_late}
\end{figure}

Because of the detection of newly forming dust, one may expect emission from CO or other dust precursor molecules. CO molecules form in SN ejecta at higher temperatures than dust grains, then efficiently cool the surrounding gas. As shown in Fig.~\ref{fig:nir_compare_early} and Fig.~\ref{fig:nir_compare_late}, we do not detect CO emission (in either the fundamental band or first overtone) in any of our NIR spectra at +71, +102, and +377~days despite dust emission being present at these epochs.

The lack of detected CO does not rule out either the presence of CO or newly forming dust. In SN~Ic~2007gr \citep{Hunter2009}, CO emission (from the first overtone) was visible at +70~days, but had disappeared by +377~days, perhaps corresponding to the optical depth decreasing such that fast electrons produced by gamma rays near the core are able to destructively interact with CO molecules in the C/O layer (although, the extremely low nucleosynthetic yield in SN~2023xgo discussed in Section~\ref{sec:discussion} makes this challenging). The presence of ionized He (which is abundant in both the ejecta and CSM of SN~2023xgo) in proximity to the CO can also accelerate its destruction without destroying dust grains if the grains are sufficiently large. Therefore, it is possible the CO was present in SN~2023xgo, but missed by our observations.

\subsection{Dust Radius}\label{sec:dustloc}
The inferred dust radius is the most direct indicator of the dust formation location when compared to the shock location at time of observation. New dust forms in the ejecta at $r \ll R_{\rm shock}$ and in the CDS at $r\leq R_{\rm shock}$, while pre-existing dust can extend to $r \gg R_{\rm shock}$. 

In our silicate dust models, the dust radius is well constrained to $r=2.3\times10^{16}$~cm. While there are no X-ray observations of SN~2023xgo from which to directly measure the shock velocity, we take the ejecta velocity as a lower limit. Estimates of the ejecta velocity from optical spectra and light curve modeling report velocities between $\sim5000-10000$~km~s$^{-1}$. Assuming a shock velocity of 10000~km~s$^{-1}$ \citep[consistent with X-ray observations of interacting SNe, e.g.][]{Immler2008, Pellegrino2024}, our estimated shock radius at +377~days post-explosion is at $3.3\times10^{16}$~cm -- external to, but roughly consistent with the silicate dust radius. This suggests that the bulk of the observed dust is forming closely behind the shock in the CDS.


However, the modeled radius is only a direct measurement of the dust radius when the dust is optically thick -- which is not the case for our carbonaceous dust models. The inferred radius in this case tends to very large radii ($r>10^{17}$~cm) and optically thin emission. Thus, we determine a lower limit of the dust radius by measuring the blackbody radius. If the dust cloud is optically thin, the blackbody radius (assuming spherical symmetry) is the smallest permitted radius which could produce the observed dust luminosity given the dust temperature.

To measure the blackbody radius, we integrate the carbonaceous dust model across all wavelengths to obtain a bolometric luminosity. We then estimate the blackbody radius as  \begin{equation}\label{eq:rbb}
r_{bb} = \sqrt{\frac{L_{bol}}{4 \pi \sigma T^4}}.
\end{equation}
We measure a blackbody radius $r_{bb} = 5.5\times$10$^{16}$~cm, which is external to the shock radius for shock velocities $\lesssim$17000~km~s$^{-1}$.

Because we cannot distinguish between dust compositions and therefore measure the radius, SED modeling alone cannot definitively determine where the emitting dust in SN~2023xgo formed. We must also mention the distinct (and perhaps most likely) possibility that there is emission from both pre-existing \textit{and} newly forming dust at distinct locations. \citet{Yamanaka2025} show an IR excess already present at +15~days, at which point the SN ejecta and CDS are still too hot for dust condensation to commence. They also find that the dust mass (measured from NIR-only observations) remains roughly constant out to $\sim$50~days. However, the spectra presented in \citet{farias2026} and this paper in Fig.~\ref{fig:attenuate} show clear increasing attenuation over time from dust interior to a velocity coordinate of $\sim$5000~km~s$^{-1}$, which cannot be pre-existing.  Qualitatively, the SED of SN~2023xgo is extremely similar to that of SN~2006jc by $\sim$+400~days, which \citet{Mattila2008} are able to describe as emission from new dust in the post-shock CSM, and pre-existing dust in the unshocked CSM.

Ultimately, the best way to disambiguate and gauge the relative contributions of pre-existing and newly forming dust in the complicated environments of interacting SNe is to obtain multiwavelength (including IR) observations over a long time baseline \citep[e.g.][]{Shahbandeh2025,Tinyanont2025, Singh2026}. A growing observed dust mass combined with the evolution of the temperature, inferred optical depth, and shock radius provides the most direct constraints on the amount of newly formed dust.

\subsection{Ongoing Circumstellar Interaction}\label{sec:interaction}

\begin{figure}
\includegraphics[width=0.44\textwidth]{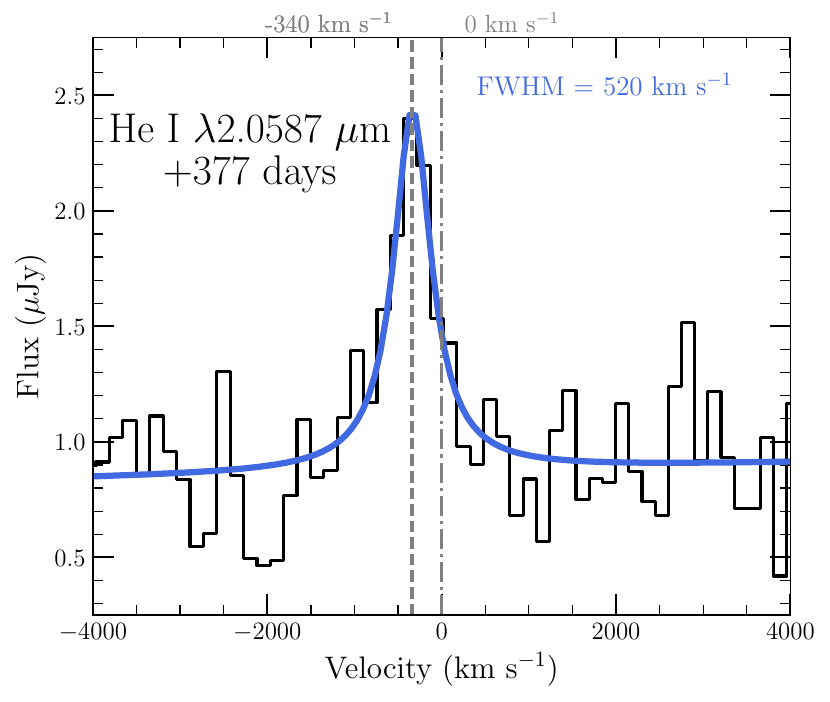}
    \caption{Detection of narrow emission from \ion{He}{1}~$\lambda$2.06~$\mu$m at +377~days post-explosion. We measure a FWHM of 520$\pm{130}$~km~s$^{-1}$ when fitting the line with a Lorentzian after subtracting the local continuum. The measured FWHM is roughly half of that measured from early-time optical spectra of SN~2023xgo \citep{Gangopadhyay2025}. The line is blueshifted by 340$\pm{40}$~km~s$^{-1}$ relative to the rest frame. This could arise from an incorrect redshift, asymmetry in the CSM, or dust attenuation of the redshifted component of the line.
    }
    \label{fig:HeLine}
\end{figure}

Shown in Fig.~\ref{fig:HeLine}, we detect narrow emission consistent with \ion{He}{1}~$\lambda$2.06~$\mu$m at +377~days post-explosion in the NIRSpec spectrum.  After subtracting the dust continuum, we fit the emission line profile with a Lorentzian and recover a full width at half maximum (FWHM) of 520$\pm{130}$~km~s$^{-1}$ and a line profile center of $-$340$\pm{40}$~km~s$^{-1}$ relative to the rest frame. Because of the low velocity relative to SN ejecta or even post-shock CSM \citep[e.g.][]{Dessart2022}, we interpret the line as emission from extended CSM external to the shock.

Both the measured FWHM of 520~km~s$^{-1}$ (perhaps set by electron scattering) and the offset of $-$340~km~s$^{-1}$ are notably lower than the CSM velocities measured from emission lines in the early-time optical spectra \citep[$\sim$1500--2000~km~s$^{-1}$,][]{Gangopadhyay2025}. One possible explanation for this discrepancy is that the CSM velocities measured at early times are systematically high relative to the progenitor outflow velocity due to radiative acceleration of the close-in CSM by the SN \citep{Wu+Fuller2022}. Similarly, the progenitor may have undergone a particularly energetic episode of mass loss immediately prior to explosion, where mass was ejected at velocities higher than the escape velocity. 
In both these cases, the late-time detection of spectral lines may provide a more constraining measurement of the progenitor outflow velocity. We discuss the implications of the lower velocity in Section~\ref{sec:discussion}.

Another possible explanation for the low velocity is that dust has formed interior to the emitting region such that it preferentially blocks emission from the far side of the SN, creating an overall narrower, and blueshifted emission line. For the narrow CSM lines to be attenuated, the dust and CSM must be oriented in a manner such that the dust emitting cross-section is comparable in area to the CSM emitting cross-section (otherwise our line-of-sight to the CSM on the far side of the SN is not significantly obstructed by dust). Two possible scenarios are (1) the dust radius and CSM emitting radius are similar (e.g. dust forming near the shock which has now traversed most of the CSM), or (2) the CSM is configured in a polar and/or collimated outflow oriented towards the observer. We note that while the emission line is roughly symmetric about $-$340~km~s$^{-1}$, the data are not high enough signal-to-noise to rule out dust attenuation of a redder component from the line profile itself.

\section{Discussion}\label{sec:discussion}
Below, we discuss the possible dust heating sources, the CSM properties in the context of potential progenitor systems, and the large dust mass associated with SN~2023xgo and implications on SN dust formation in the early universe.


\subsection{Bolometric Luminosity and Heating Source}\label{sec:bolo}

Because the heat capacity of dust grains is very small \citep{Draine+2001}, the dust emission at +377~days requires an ongoing heating source to power the dust luminosity. By extrapolating and integrating our modeled dust SEDs across all wavelengths (and assuming there are no colder components hidden at larger wavelengths), we measure a bolometric luminosity of $\sim$1$\times10^{40}$~erg~s$^{-1}$ at +377~days. $\gtrsim$90~\% of the luminosity is emitted in wavelengths captured by the \textit{JWST} observations. 

Below, we discuss the feasibilities of (1) radiative heating from ongoing circumstellar interaction, (2) radiative heating from radioactive decay close to the dust, (3) heating from an infrared echo of the peak SN luminosity, and lastly (4) collisional heating from the SN shock.

First, the detection of narrow \ion{He}{1} confirms that circumstellar interaction persists at +377~days, and therefore contributes at least \textit{some} of the luminosity heating the dust. Although the light curve models presented in \citet{Gangopadhyay2025} and \citet{farias2026} predict the interaction luminosity to drop off quickly after $\gtrsim$30~days, the detection of narrow He directly indicates the SN shock continues to encounter extended CSM not probed by the early-time data. The luminosities of SNe~Ibn/Icn at $>$+1~year are poorly sampled. We note that SNe~IIn commonly show late-time optical luminosities of $L_{\rm +1~year}\gtrsim10^{40}$~erg~s$^{-1}$ \citep[e.g.,][]{Fox2013}, although as a class SNe~IIn are more luminous and slowly evolving than SNe~Ibn/Icn \citep{Hosseinzadeh2017, Ransome2025}. Nevertheless, because of the direct detection of narrow \ion{He}{1}, we consider interaction-heating feasible.

Next, we consider radiative heating from radioactive decay in the SN ejecta. \citet{Gangopadhyay2025} and \citet{farias2026} place upper limits on the amount of radioactive $^{56}$Ni synthesized in SN~2023xgo at $\leq$0.04~M$_{\odot}$ and $\leq$0.005~M$_{\odot}$, respectively from modeling the early-time light curve. While we note that significant degeneracies exist in the light curve modeling,  $\leq$0.04~M$_{\odot}$ $^{56}$Ni is consistent with that found in other SNe~Ibn/Icn \citep{Pellegrino2022, farias2026}. Using an Arnett radioactive decay model \citep{Arnett1982} and optical and gamma-ray opacities consistent with \citet{Gangopadhyay2025}, we find a luminosity of $\sim10^{39}$~erg~s$^{-1}$ at +377~days. Assuming a large fraction of the luminosity heats the observed dust, radioactive decay can plausibly account for a meaningful fraction, but not the majority of the heating.

Next, we investigate the feasibility of an infrared echo of the peak SN luminosity wherein the late-time emission arises from time delays associated with the light paths the SN emission takes to different regions of the CSM before being re-emitted towards the observer. We follow the prescription laid out in \citet{Fox2011} to calculate expected echo-heated dust temperatures given a peak SN luminosity as a function of the echo radius. We take the peak SN bolometric luminosity of $\sim$10$^{43}$~erg~s$^{-1}$ measured in \citet{Gangopadhyay2025}. 

\begin{figure}
\includegraphics[width=0.48\textwidth]{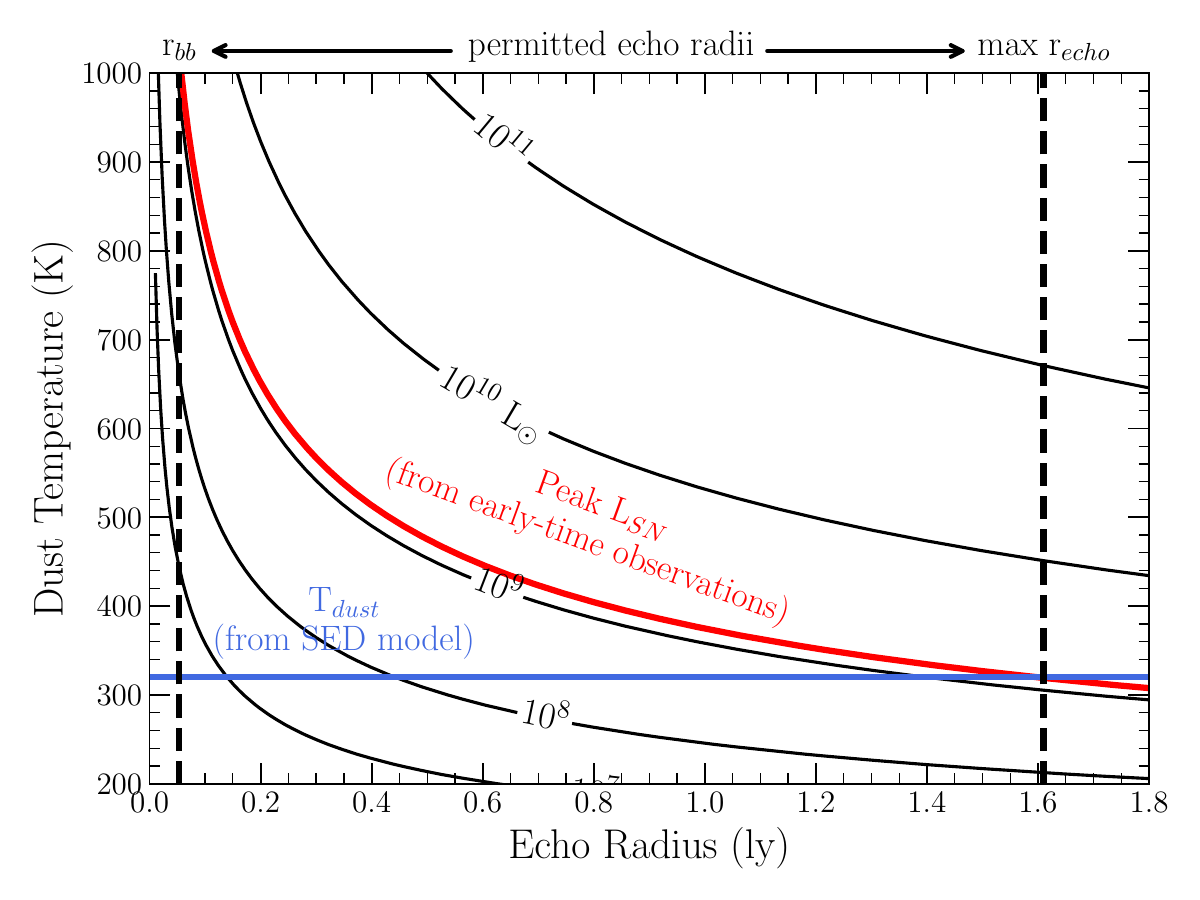}

    \caption{Carbonaceous dust temperature at different echo radii given a peak SN luminosity adapted from \citep{Fox2011}. Contours show different peak SN luminosities in units of L$_{\odot}$, where the parameter space below each represents permitted dust/echo parameters given the input energy. The peak luminosity of SN~2023xgo from \citet{Gangopadhyay2025} is marked in red. The input luminosity is treated as a single pulse for a given echo radius; see details in \citet{Fox2011}. An IR echo is energetically permitted between 0.05~ly (4.7$\times10^{16}$~cm, the blackbody radius from our SED fits) and 1.6~ly (1.5$\times10^{17}$~cm). Our silicate dust models cannot be heated by an IR echo. 
    }
    \label{fig:echo_energy}
\end{figure}

We show the allowed dust temperatures for a range of peak luminosities in Fig.~\ref{fig:echo_energy}. For the modeled (mass-weighted average) carbonaceous dust temperature of 320~K, echo heating is energetically permitted for a dust radius $<$1.6~lyr (well above our lower limit of $r_{bb}=5\times10^{16}$~cm). Silicate dust grains require more energy to heat to the same temperature, and thus the upper limit of the permitted echo radius is $\sim$0.12~lyr. However, the directly fit radius of the silicate dust sphere is $r\approx2.5\times10^{16}$~cm, or 7~light days, which cannot manifest in an IR echo of the peak SN luminosity at +377~days. A rebrightening of the SN at +1~year to the luminosities required to heat the dust is extremely unlikely. Thus, an alternative heating source is required if we assume the silicate composition.

Next, we investigate the feasibility of collisional heating of pre-existing dust from the SN shock. We employ the equations from \citet{Fox2010} to estimate the volume of gas processed by the forward shock over the grain sputtering timescale. Then, assuming a dust-to-gas mass ratio of 0.01, the dust mass is
\begin{equation}\label{eq:collision}
M_{\mathrm{dust}} \approx 0.0028 \left(\frac{\nu_\mathrm{s}}{ \mathrm{15,000~km\ s}^{-1}}\right)^3 \left(\frac{t}{\mathrm{year}}\right)^2 \left(\frac{a}{\mu \mathrm{m}}\right)~{M_{\odot}},
\end{equation}
where $\nu_\mathrm{s}$ is the shock velocity, $t$ is the time since explosion, and $a$ is the dust grain size. Following \citet{Zsiros2024} and \citet{Pearson2025}, we determine a reasonable range of dust masses by examining velocity and grain size ranges of  $v_{\mathrm{shock}}= $ 5000 -- 15000~km~s$^{-1}$ and $a = 0.005$ -- 0.1~$\mu$m.

 We achieve a range of collisionally heated dust masses between 5$\times$10$^{-6}$~M$_{\odot}$ and 3$\times$10$^{-4}$~M$_{\odot}$, which cannot explain the observed dust masses. The mass is insufficient even before accounting for the dust-free cavity interior to the evaporation radius. While a more aggressive dust-to-gas mass ratio could help alleviate these tensions, the highest ratio found in the \citet{Lau2020} sample of colliding wind WR binaries was only $\sim$0.15, with most falling far below. We therefore consider collisional heating unlikely.

Thus, while the precise heating source of the dust associated with SN~2023xgo is uncertain, we have constrained the possibilities depending on the inferred composition. Carbonaceous dust at large radii can be reasonably explained as an IR echo of the peak SN luminosity. Silicate dust, which must be confined to a radius of $r\approx2.5\times10^{16}$~cm to explain the observed SED, cannot be explained by an IR echo of the peak SN luminosity. Instead, we find heating by a combination of radioactive decay and ongoing circumstellar interaction to be plausible for both compositions.

\subsection{CSM Properties and Connections to Progenitors}

The velocity evolution and detection of CSM interaction at +377~days add to a narrative that the mass-loss history of SN~2023xgo is complex and likely varying with time. \citet{Gangopadhyay2025} infer a radially variable CSM density profile by combining early light curve and late-time (through +63~days) spectral modeling. They estimate a low density (mass loss rate of 10$^{-3}$–-10$^{-4}$~M$_{\odot}$~yr$^{-1}$) component extending to $\lesssim10^{13}$~cm, and a \textit{more} dense (0.1--2.7~M$_{\odot}$~yr$^{-1}$) component extending out to $\sim10^{15}$~cm. In our spectrum taken at +377~days, the shock even still continues to encounter CSM at $>10^{16}$~cm, corresponding to material lost $\sim$10~years prior to explosion.

The bulk velocity offset ($-340$~km~s$^{-1}$) and line width ($\rm{FWHM}=540$~km~s$^{-1}$) of \ion{He}{1} at +377~days indicate a CSM outflow that is substantially lower velocity than inferred from early-time measurements. \citet{Davis2023} discuss a similar low-velocity outflow in the SN~Icn~2022ann, and find the outflow to be inconsistent with typical WR escape velocities ($\gtrsim$ 1000~km~s$^{-1}$). If this line presents a true limit on the outflow velocity, it very strongly rules out single WC/WO scenarios, whereas several configurations of more moderate mass binaries can reproduce this outflow velocity \citep[e.g.][]{Tsuna2024a}. \citet{Wu+Fuller2022} present the possibility that measuring the CSM velocities from early-time spectra will yield systematically high estimates of the outflow velocity due to radiative acceleration of the close-in CSM by the SN.




Next, we explore the global CSM properties under the assumption that the bulk of the dust associated with SN~2023xgo is pre-existing. \citet{Yamanaka2025} measure a dust mass of $\sim$10$^{-4}$~M$_{\odot}$ from NIR observations, and employ a standard dust-to-gas mass ratio of 0.01 to infer a CSM mass of $\sim$10$^{-2}$~M$_{\odot}$. We measure a much larger dust mass of $\sim$10$^{-2}$~M$_{\odot}$ from our \textit{JWST} data, from which we infer a CSM mass of $\sim$1~M$_{\odot}$ assuming the same dust-to-gas mass ratio. This more massive estimate is consistent with the CSM masses estimated in SNe~Ibn/Icn from optical light curve modeling \citep[e.g.,][]{Pellegrino2022, farias2026}. 

However, we caution that the dust-to-gas mass ratio in binary systems is highly sensitive to the precise physical conditions (e.g., \cite{Lau2020} find that the circumstellar dust-to-gas mass ratio in colliding-wind WR binaries varies by 6 orders of magnitude). At the very least, the dust mass itself (if pre-existing) serves as a concrete lower limit of the CSM mass, whereas estimates derived from light curve modeling depend on several model assumptions.

The dust SED itself in SN~2023xgo is qualitatively similar to that of dust around carbon-rich WR (WC) stars. Notably, a binary companion is believed to be necessary to form dust in the fast and hot circumstellar environments of WCs \citep{Usov1991}.  Estimates of WC dust formation rates measured in \cite{Lau2020} integrated over a typical WC lifetime of $\sim$10$^5$~yr could result in comparable overall dust masses. In comparison, dust masses formed by binary mergers such as mass ejections in luminous red novae are more modest ($\sim$10$^{-3}$~M$_{\odot}$), though the extrapolation of these dust masses to their final values is uncertain \citep{Karambelkar2026}, and the dust produced in other binary interactions is an ongoing area of study.

While the exact progenitor of SN~2023xgo remains unclear, we have demonstrated that \textit{JWST} observations can place unique constraints on the CSM properties of SNe~Ibn/Icn. In the future, combining a full complement of multi-wavelength observations and pre-explosion monitoring (i.e., from LSST and \textit{Roman}) will paint a full picture of SN Ibn/Icn mass-loss histories, their system geometries, and whether or not they are explosions of single massive stars \citep[e.g.][]{Immler2008, Dessart2022, Lu2023, Baer-Way2025}.

\subsection{Implications for Early-Universe Dust}

\begin{figure*}
\centering
\includegraphics[width=0.95\textwidth]
{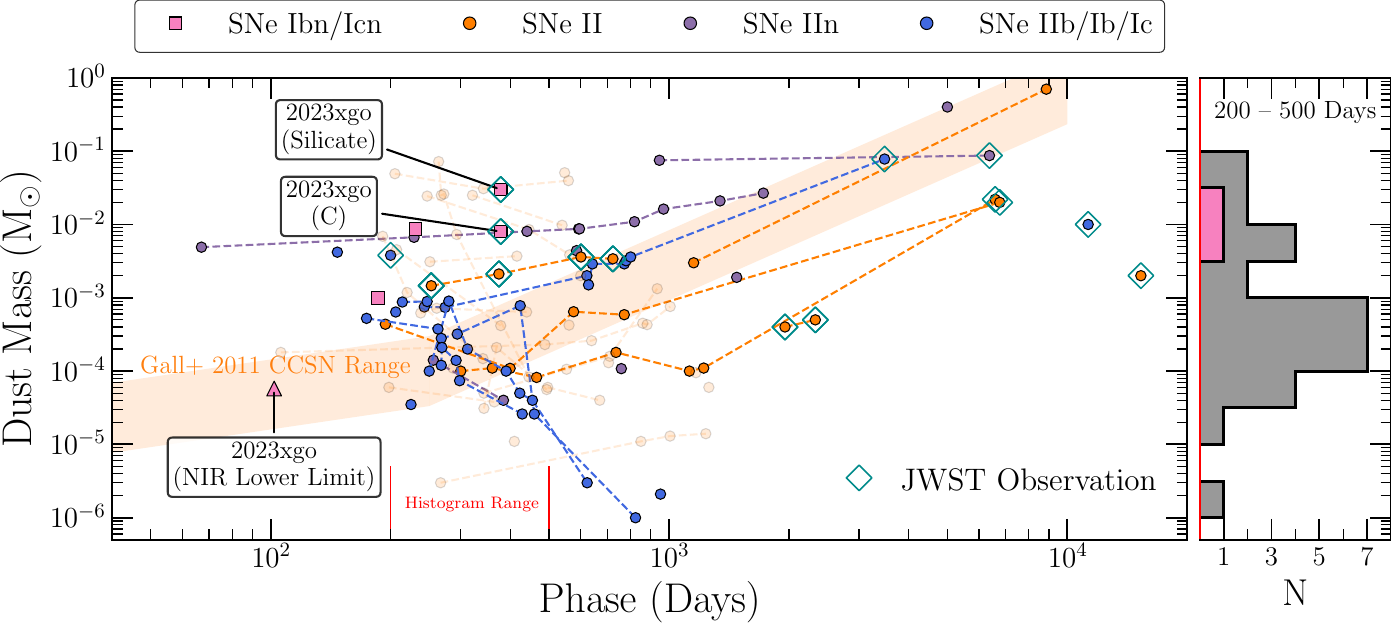}
\label{fig:dustphase}
    \caption{\textbf{Left:} Compilation of observed dust masses versus phase for SNe~Ibn/Icn \citep[][this paper]{Mattila2008, Gan2021} compared to several types of CCSNe \citep[][and references therein]{Meikle2007, Gall2011, Wesson2015, Gallagher2012, Shahbandeh2023, Hosseinzadeh2023, Zsiros2024, Shahbandeh2025, Tinyanont2025, Pearson2025, Singh2026}. Dust mass measurements derived from \textit{JWST} observations are outlined with teal diamonds. SNe~II with no \textit{JWST} or \textit{Herschel} observations are plotted in a lower opacity. We present the total observed dust mass without differentiating between newly forming and pre-existing dust.  We do not plot any non-detections and note that several of the dust masses are derived from \textit{Spitzer} observations in only 2 filters \citep{Szalai2019}. Given these systematic limitations, we caution against interpreting these data as a true distribution of SN dust masses. \textbf{Right:} Histogram of observed dust masses derived from observations between 200 and 500 days post-explosion. We include the single highest dust mass measurement for each SN within this phase range. SNe~Ibn/Icn are shown in pink, while all other CCSNe are shown in gray. SNe~Ibn/Icn are among the dustiest SNe observed in this parameter space.}
\end{figure*}

Shown in Fig.~\ref{fig:attenuate}, new dust formation in SN~2023xgo commences by +67~days, and the dust mass observed in SN~2023xgo (and SN~2006jc) at +1~year is substantially higher than other CCSNe observed at a similar phase (Fig.~\ref{fig:dustphase}). Whether the large dust mass is predominantly pre-existing or newly forming, it is clear that the dense and C/O-rich circumstellar environment around SN~2023xgo has facilitated substantial dust formation. 

Recent \textit{JWST} observations of interacting SNe of several types have confirmed that they form \textit{new} dust post-explosion in quantities and rates that surpass their non-interacting counterparts \citep[e.g.][]{Shahbandeh2025, Tinyanont2025, Singh2026}. The clear takeaway is that the presence of dense CSM (and therefore pre-SN mass loss) boosts the production of dust both before and after the terminal SN explosion.

\cite{Sarangi2022a} test the impact of mass-loss rate and ejecta mass on the speed and final mass of dust formation in H-rich interacting SNe. In their most rapidly dust-forming case (low ejecta mass of 6~M$_{\odot}$ and high pre-explosion mass-loss rate of 10$^{-1}$~M$_{\odot}$~yr$^{-1}$), dust still did not begin forming until 620~days post-explosion. Given that there is unambiguously newly formed dust at $<$100~days in several interacting SNe (SNe~2005ip, 2006jc, 2010jl, 2023xgo), further investigation into exactly how the prompt dust formation occurs is required.

There are several pathways for interacting SNe (including SNe~Ibn/Icn) to produce a meaningful fraction of the high-\textit{z} SN dust budget despite their rarity at low \textit{z}. While lower metallicities will suppress wind-driven mass loss, a top-heavy initial mass function \citep{Marks2012} and higher binary interaction rates \citep{Sana2025} may directly boost the rates of SN~Ibn/Icn-like transients at high \textit{z}. Furthermore, CDS dust formation is typically thought to rapidly produce large dust grains which survive dispersal into the ISM at far higher rates than smaller grains \citep[e.g.,][]{Nozawa2007, Gall2011}.

 Compounding the potential higher rates in the early universe with an increased chance for the dust they produce to actually survive in the ISM indicates that SNe~Ibn/Icn (and similar interacting SNe) could be overrepresented in the SN dust budget relative to their local rates. Binary mass transfer and envelope-stripping at lower metallicities are extremely active areas of ongoing research \citep[e.g.,][]{Doughty2021, Andrews2025}, and a better understanding of rates will be key ingredients in a fully balanced SN dust budget.

Even if the dust in SN~2023xgo is entirely pre-existing in the CSM, it then serves as a benchmark for the amount of dust that high-\textit{z} analogs to its progenitor could produce over their lifetimes. \cite{Lau2020} use binary population synthesis models to show that dust produced in Wolf-Rayet binaries and RSGs dominate the dust budget of star-forming galaxies for the first $\sim$100~Myr. While low metallicities will suppress the formation of single WRs, binary interactions again present a viable channel for their creation (and subsequent dust production) in the early universe.

\section{Conclusion}\label{sec:conclusion}
We have presented near- and mid-infrared observations of the Type Ibn/Icn SN~2023xgo spanning +71 to +377~days post-explosion, including \textit{JWST} spectroscopic and photometric observations reaching out to $\sim$23~$\mu$m. Owing to their rapid photometric evolutions in UV and optical wavelengths, SNe~Ibn/Icn are rarely observed in the infrared. Comparable data has not been reported for any SN~Ibn since SN~2006jc.

By $\sim$+101~days post-explosion, the SED of SN~2023xgo is dominated by dust emission. We place a lower limit on the dust mass at +101~days of 6.8$\times$10$^{-5}$~M$_{\odot}$ from observations out to $\sim$5~$\mu$m. At +377~days, we find that our modeling still cannot distinguish between silicate and carbonaceous dust compositions. In the carbonaceous dust case, we recover a total dust mass of 8.1$\times$10$^{-3}$~M$_{\odot}$, with temperature components of 315 and 526~K. The emission is optically thin, and thus we refer to the blackbody radius of $5.5\times10^{16}$~cm as a lower limit for the dust sphere radius. In the silicate dust case, the emission is optically thick and therefore the mass can be arbitrarily large. We report the lower 2.5th percentile of our mass posterior distribution as an estimate of the lower limit, yielding a total of 2.8$\times$10$^{-2}$~M$_{\odot}$ at temperature components of 385 and 608~K. In this case the radius is well constrained to $2.3\times10^{16}$~cm.

The emission line profiles from optical and NIR spectra out to $\sim$70~days show a progressive blueshifting from dust formation interior to the line-emitting region -- a telltale sign of active dust formation. Emission from molecular gas (e.g. CO and SiO) is not seen in any of our spectra, although we do not interpret this as the lack of any molecule formation.

Lastly, we detect narrow \ion{He}{1} emission in the \textit{JWST} spectrum, which we interpret as ongoing interaction between the shock and extended circumstellar material. The emission is narrow (FWHM~$\approx520$~km~s$^{-1}$) and blueshifted by 340~km~s$^{-1}$, indicating a lower velocity outflow than that probed by the early time spectra. Unless there is substantial deceleration from interactions with the nearby ISM, this low velocity cannot be explained by a wind from a single WR-like star. We find ongoing circumstellar interaction and radioactive decay to be feasible heating mechanisms for the dust, although an IR echo of the peak SN luminosity is also feasible for carbonaceous dust grains.

This dust mass is among the highest observed in any SN at a similar phase. As a class, SNe~Ibn/Icn posses very high dust masses, indicating that their unique circumstellar environments facilitate the formation of copious amounts of dust (whether before or after the SN explosion). Future observations will be critical in disambiguating pre-existing and newly forming dust and charting out the full mass-loss histories of these extreme SNe. 

\section{Acknowledgements}

The UCSC transients team is supported in part by STScI grant JWST-DD-6838 and by a fellowship from the David and Lucile Packard Foundation to R.J.F.
D.F.\ is supported by a VILLUM FONDEN Young Investigator Grant (project number 25501) a Villum Experiment grant (VIL69896) and by research grants (VIL16599, VIL54489) from VILLUM FONDEN.
C.D.K.\ gratefully acknowledges support from the NSF through AST-2432037, the HST Guest Observer Program through HST-SNAP-17070 and HST-GO-17706, and from JWST Archival Research through JWST-AR-6241 and JWST-AR-5441. 
E.R.R.\ acknowledges the Heising-Simons Foundation and NSF: AST--1852393, AST--2150255, and AST--2206243.
Q.W.\ is supported by the Sagol Weizmann-MIT Bridge Program.

This work is supported by the National Science Foundation under Cooperative Agreement PHY-2019786 (The NSF AI Institute for Artificial Intelligence and Fundamental Interactions, http://iaifi.org/).
Parts of this research were supported by the Australian Research Council Centre of Excellence for Gravitational Wave Discovery (OzGrav), through project number CE230100016.

This work is based in part on observations made with the NASA/ESA/CSA James Webb Space Telescope. The data were obtained from the Mikulski Archive for Space Telescopes at the Space Telescope Science Institute, which is operated by the Association of Universities for Research in Astronomy, Inc., under NASA contract NAS 5--03127 for JWST. These observations are associated with program 6838 (PI Davis).  Support for program 6838 was provided by NASA through a grant from the Space Telescope Science Institute, which is operated by the Association of Universities for Research in Astronomy, Inc., under NASA contract NAS 5-03127. 

This work is based in part on observations obtained at the international Gemini Observatory, a program of NSF NOIRLab, which is managed by the Association of Universities for Research in Astronomy (AURA) under a cooperative agreement with the U.S.\ National Science Foundation on behalf of the Gemini Observatory partnership: the U.S.\ National Science Foundation (United States), National Research Council (Canada), Agencia Nacional de Investigaci\'{o}n y Desarrollo (Chile), Ministerio de Ciencia, Tecnolog\'{i}a e Innovaci\'{o}n (Argentina), Minist\'{e}rio da Ci\^{e}ncia, Tecnologia, Inova\c{c}\~{o}es e Comunica\c{c}\~{o}es (Brazil), and Korea Astronomy and Space Science Institute (Republic of Korea). This work was enabled by observations made from the Gemini North telescope. The scientific community is honored to have the opportunity to conduct astronomical research on Maunakea in Hawai‘i. We recognize and acknowledge the very significant cultural role and reverence of Maunakea to the Kanaka Maoli (Native Hawaiians) community.

\bibliography{2023xgo.bib}{}
\bibliographystyle{aasjournal}

\appendix
\label{appendix}


\setcounter{figure}{0}                       
\renewcommand\thefigure{A.\arabic{figure}}

\begin{table}[h]
\centering
 \caption{Dust SED modeling priors. Priors are uniform within the reported ranges.} \label{tab:dustpriors}
 \hspace*{-1.6cm}
 \begin{tabular}{ l c c c c c c}
    \hline
     Model & $\log M_{1}$ & $\log M_{2}$ & $T_{1}$ & $T_{2}$ & $\log R_{\rm in}$ & $\log R_{\rm out}$ \\
      & [M$_{\odot}$] & [M$_{\odot}$] & [K] & [K] & [cm] & [cm] \\
     \hline
     Si+Si sphere & $[-4,0]$ & $[-6,-2]$ & $[100,500]$ & $[400,1800]$ & -- & $[15,19]$\\
     C+C sphere & $[-4,0]$ & $[-6,-2]$ & $[150,500]$ & $[500,1800]$ & -- & $[15,19]$\\
     Si+Si shell & $[-4,-1]$ & $[-6,-2]$ & $[150,500]$ & $[500,1800]$ & $[10,17]$ & $[15,19]$\\
     C+C shell & $[-4,-1]$ & $[-6,-2]$ & $[150,500]$ & $[500,1800]$ & $[10,17]$ & $[15,19]$\\
     C+Si sphere & $[-4,0]$ & $[-6,0]$ & $[100,500]$ & $[400,1800]$ & -- & $[15,19]$\\
     C+C optthick & $[-4,-1]$ & $[-6,-2]$ & $[150,500]$ & $[500,1800]$ & -- & $[15,17]$\\
    \hline
 \end{tabular}
\end{table}

\begin{figure*}[!ht]
        \begin{center}
        \includegraphics[width=0.48\textwidth]{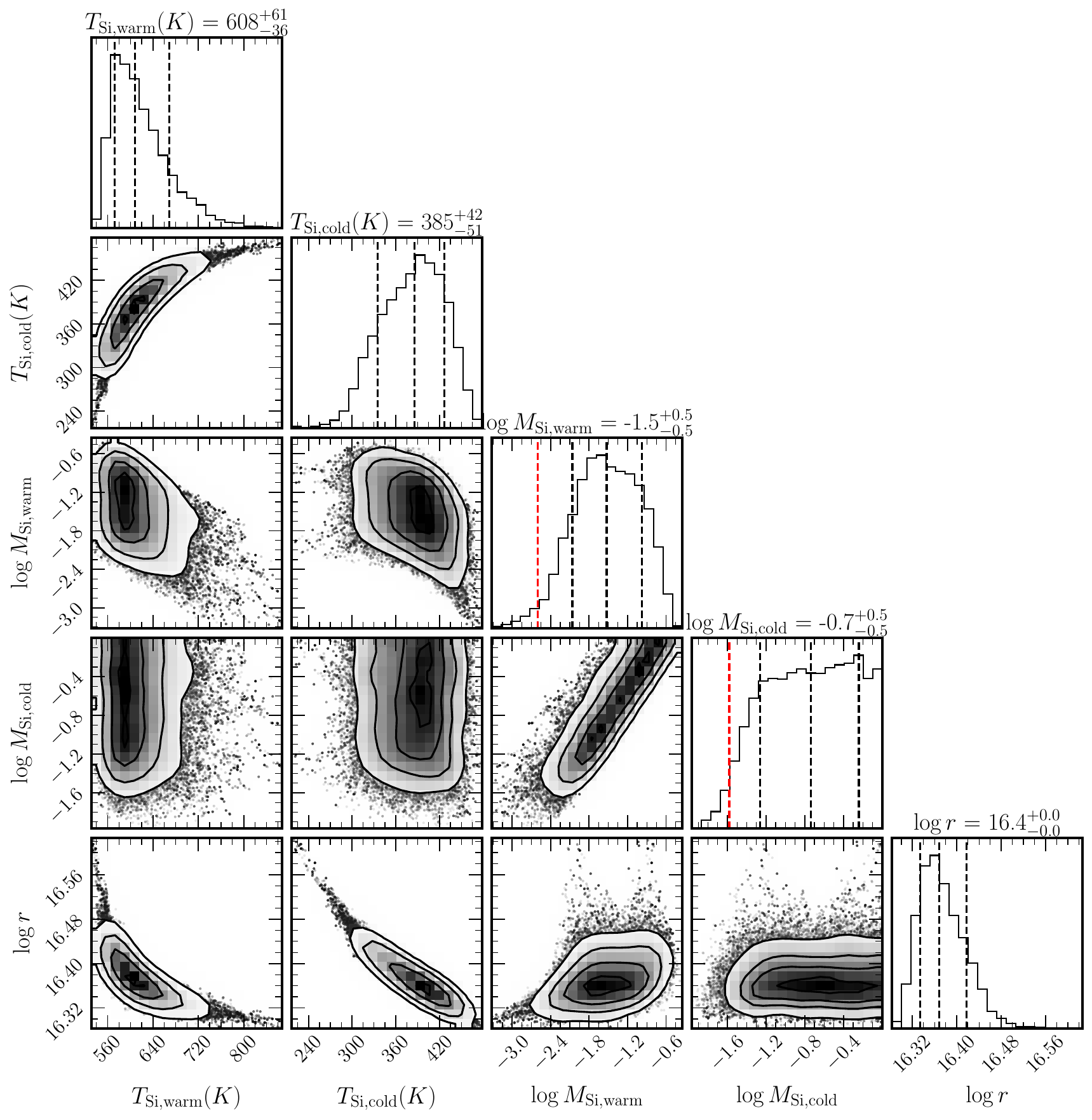}
        \hspace{-0.1cm}
        \includegraphics[width=0.48\textwidth]{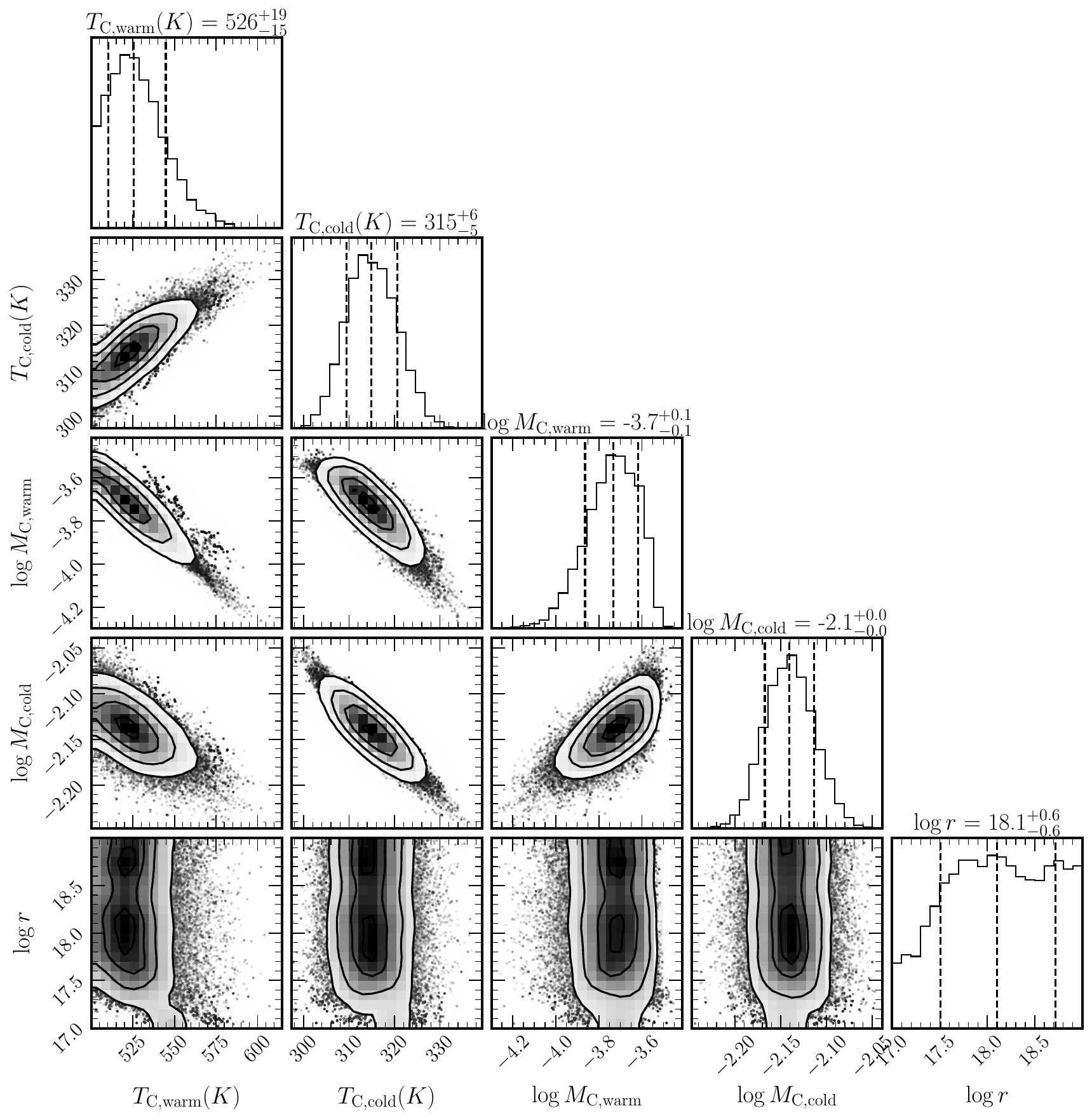}
        \end{center}
        \caption{Corner plots showing posterior distributions for our dust SED modeling. We show the \texttt{Si+Si sphere} (left) and \texttt{C+C sphere} (right) models, which both assume uniform spherical geometry. In the \texttt{Si+Si sphere} model, the emission is optically thick and the mass can tend to arbitrarily high values without changing the observed SED. We adopt the 2.5th-percentile value as a lower limit for the mass (shown in red). For \texttt{C+C sphere}, the emission is optically thin and therefore the radius can tend to arbitrarily high values. We adopt the blackbody radius $r_{bb}=5.5\times10^{16}$~cm as a lower limit.
        }

        \label{fig:corner}
\end{figure*}

\begin{table*}[h]
\centering
 \caption{Log of Photometric Observations} \label{tab:photlog}
 \hspace*{-1.6cm}
 \begin{tabular}{ l c c c c}
    \hline
     Filter & MJD & Phase\tablenotemark{a} \ & Flux Density & Total Integration \\
     & & [days]& [mJy] & [s]\\
     \hline
     F560W\tablenotemark{b} & 60639.40 & 377 &$0.0509\pm 0.0044$ & 89 \\
    F1280W & 60639.46 & 377 & {$0.0930\pm0.0015$} & 355 \\
    F1500W & 60639.47 & 377 & {$0.0662\pm0.0024$} & 444 \\
    F1800W & 60639.48 & 377 & {$0.0822\pm0.0024$} & 921 \\
    F2100W & 60639.50 & 377 & {$0.0752\pm0.0044$} & 2631 \\
    W1 & 60360\tablenotemark{c} & 101 & {$0.2105\pm0.0100$} & 223 \\
    W2 & 60360\tablenotemark{c} & 101 & {$0.2414\pm0.0250$} & 223 \\

    \hline
 \end{tabular}

 \tablenotetext{a}{Rest-frame days since explosion on 60257.22 MJD \citep{Gangopadhyay2025}}
 \tablenotetext{b}{MIRI LRS acquisition image}
 \tablenotetext{c}{Averaged over 29 visits}
\end{table*}

\begin{table}[H]
\centering
 \caption{Log of Spectroscopic Observations} \label{tab:speclog}
 \hspace*{-1.6cm}
 \begin{tabular}{ l c c c c c c}
    \hline
     Instrument & Disperser & MJD & Phase\tablenotemark{a} & Spectral Resolution & Wavelength Range & Total Integration \\
      & &  & [days]& [R] & [$\mu$m] & [s]\\
     \hline
     Gemini-N/GNIRS & 32 l/mm & 60263 & +6 & 1800 & 0.9--2.5 & 1500 \\
     Gemini-N/GNIRS & 32 l/mm & 60329.23 & +71 & 1800 &0.9--2.5  & 5040 \\
     Gemini-N/GNIRS & 32 l/mm & 60360.22 & +102 & 1800 &0.9--2.5  & 5040 \\
     \textit{JWST}/NIRSpec & G235M & 60639.32 & +377 & 1000 & 1.6--3.2 & 4551 \\
     \textit{JWST}/NIRSpec & G395M & 60639.37 & +377 & 1000 & 2.9--5.3 & 2451  \\
     \textit{JWST}/MIRI & LRS double prism &  60639.43 & +377 & 100 & 5.0--14 & 3591 \\

    \hline
 \end{tabular}
\tablenotetext{a}{Rest-frame days since explosion on 60257.22 MJD \citep{Gangopadhyay2025}}
\end{table}

\end{document}